\documentclass[11pt]{article}
\usepackage{amsfonts}
\usepackage{amssymb}
\usepackage{latexsym}
\usepackage{graphicx}
\usepackage[english]{babel}
\begin{document}
\input epsf

\def\thetat{{\tilde\theta}}
\def\psit{{\tilde\psi}}
\def\phit{{\tilde\phi}}
\def\rb{{\bar r}}
\def\thetab{{\bar \theta}}
\def\psib{{\bar \psi}}
\def\phib{{\bar \phi}}
\def\Qb{{\bar Q}}
\def\yh{\hat{y}}
\def\k{\kappa}
\def\sigmah{\hat\sigma}

\def\nn{\nonumber\\}
\def\t{\tilde}
\def\p{\partial}
\def\h{{1\over 2}}
\def\be{\begin{equation}}
\def\bea{\begin{eqnarray}}
\def\ee{\end{equation}}
\def\eea{\end{eqnarray}}
\def\d{\partial}
\def\la{\lambda}
\def\eps{\epsilon}
\def\bb{\bigskip}
\def\mm{\medskip}
\newcommand{\dm}{\begin{displaymath}}
\newcommand{\edm}{\end{displaymath}}
\renewcommand{\b}{\tilde{B}}
\newcommand{\gm}{\Gamma}
\newcommand{\ac}[2]{\ensuremath{\{ #1, #2 \}}}
\renewcommand{\ell}{l}
\newcommand{\z}{\ell}
\newcommand{\newsection}[1]{\section{#1} \setcounter{equation}{0}}
\def\bb{$\bullet$}

\def\q{\quad}

\def\bn{B_\circ}

\let\a=\alpha \let\b=\beta \let\g=\gamma \let\d=\delta \let\e=\epsilon
\let\c=\chi \let\th=\theta  \let\k=\kappa
\let\l=\lambda \let\m=\mu \let\n=\nu \let\x=\xi \let\r=\rho
\let\s=\sigma \let\t=\tau
\let\vp=\varphi \let\vep=\varepsilon
\let\w=\omega      \let\G=\Gamma \let\D=\Delta \let\Th=\Theta
                      \let\P=\Pi \let\S=\Sigma

\def\h{{1\over 2}}
\def\t{\tilde}
\def\r{\rightarrow}
\def\nn{\nonumber\\}
\let\bm=\bibitem
\def\Kt{{\tilde K}}

\let\p=\partial

\begin{flushright}
%OHSTPY-HEP-T-03-012\\
\end{flushright}
\vspace{20mm}
\begin{center}
{\LARGE  A microstate  for the 3-charge black ring}
\\
\vspace{18mm}
{\bf   Stefano Giusto\footnote{giusto@mps.ohio-state.edu}, Samir D.  
Mathur\footnote{mathur@mps.ohio-state.edu}
and Yogesh K. Srivastava\footnote{yogesh@mps.ohio-state.edu}}\\

\vspace{8mm}
Department of Physics,\\ The Ohio State University,\\ Columbus,
OH 43210, USA\\
\vspace{4mm}
\end{center}
\vspace{10mm}
\thispagestyle{empty}
\begin{abstract}

We start with a 2-charge D1-D5 BPS geometry that has the shape of a  
ring; this geometry is regular everywhere. In the dual CFT there exists  
a perturbation that creates one unit of excitation for left movers, and  
thus adds one unit of momentum P.
This implies that there exists a corresponding normalizable  
perturbation on the near-ring D1-D5 geometry. We find this  
perturbation, and observe that it is smooth everywhere. We thus find an  
example of `hair' for the black ring carrying three charges -- D1, D5  
and one unit of P. The near-ring geometry of the D1-D5 supertube can be  
dualized to a D6 brane carrying fluxes corresponding to the `true'  
charges, while the quantum of P dualizes to a D0 brane. We observe that  
the fluxes on the D6 brane are at the threshold between bound and  
unbound states of D0-D6, and our wavefunction helps us learn something  
about binding at this threshold.

\end{abstract}
\newpage
\setcounter{page}{1}
\renewcommand{\theequation}{\arabic{section}.\arabic{equation}}
\section{Introduction}\label{intr}
\setcounter{equation}{0}

The classical geometry of a black hole has `empty space' near the  
horizon. Pair creation in this region leads to the information paradox.  
String theory suggests that the black hole interior is
not the naive classical one; rather the information of the state of the  
hole is distributed throughout the  interior of the horizon
(for a review of some basic ideas  see \cite{fuzzball}).
Such a picture holds for all 2-charge extremal states, and has been  
shown to also extend for a subset of 3-charge states.

In 4+1 dimensions we can have not only black holes but also black rings  
\cite{emparan}.  We would therefore like to construct microstates for  
the ring. The goal of this paper is to construct a simple 3-charge  
extremal state for the ring, where we start with a  ring carrying two  
charges D1,D5 and add
a wave carrying one unit of momentum P, the third charge.

There has been a lot of recent progress on black rings.  The entropy  
for the ring can be obtained by computing it for a short straight  
segment of a ring and multiplying by the total length of the ring  
\cite{bkmicro}. A subset of 3-charge rings can be obtained as  
supertubes made out of branes \cite{tubering}.

We are interested in the gravity description of microstates.  In  
\cite{mss,3charge} dual geometries were found for a discrete subset of  
CFT states. But even though these states have a large angular momentum,  
they do not look like `rings', since we cannot find a sphere $S^2$ that  
will surround a `ring' shaped interior. In \cite{bw2,gim} a method was  
developed to find large families of 3-charge BPS states, in terms of  
the choice of locations of poles of certain harmonic functions  
appearing in the metric \cite{basicmethod}. While the CFT states for  
these geometries are not known, it was argued that these geometries  
represented bound states
because there is a nonzero flux on spheres $S^2$ linking the poles. Assuming  
that this argument is correct we can make geometries that are like  
rings, and that have no horizon and no singularity.

In the present paper we would like to construct a gravity description  
of microstates in a case where we also know the dual CFT state. Thus in  
gravity terms our construction will be more modest than the ones  
obtainable from
\cite{bw2,gim}; the third charge will be only a small perturbation on  
our 2-charge ring. On the other hand since we will know the microscopic  
origin of the state, we are assured that we have a bound state and we  
can also develop some intuition for how CFT operations act in the  
gravity picture.

In spirit our computation is similar to the computation in \cite{mss},  
where one unit of momentum was added to a D1-D5 extremal state. We will  
again take the same D1-D5 state, and add a unit of momentum using the  
same fields, but will be working in a very different limit from the one  
used in \cite{mss}. In \cite{mss} we had chosen our moduli so that the  
D1-D5 geometry had a large AdS type region, which went over to flat  
space at infinity. This geometry is depicted in Fig.1(a). The  
wavefunction of the quantum carrying the momentum is peaked in the AdS  
region, falling off at infinity in a normalizable way. By contrast in  
the present paper we will take a limit of the moduli so that the D1-D5  
state looks like a thin ring, depicted in Fig.1(b). Consider a  
short segment of this ring, which looks like a straight tube  
(Fig.1(c)). The bound state wavefunction must now appear as a  
wavefunction localized in the vicinity of this tube, falling  to  
zero away from this tube, and regular everywhere inside. We find this  
wavefunction, thus obtaining a simple but explicit example of `hair'  
for the black ring. The fact that the
wavefunction is regular everywhere suggests that no horizon or  
singularity should form even for a non-infinitesimal  
deformation,  so the result supports a `fuzzball' picture for the  
black ring.

\begin{figure}[htbp]
\begin{center}
\vskip -2.0truecm
\includegraphics[width=4.5in]{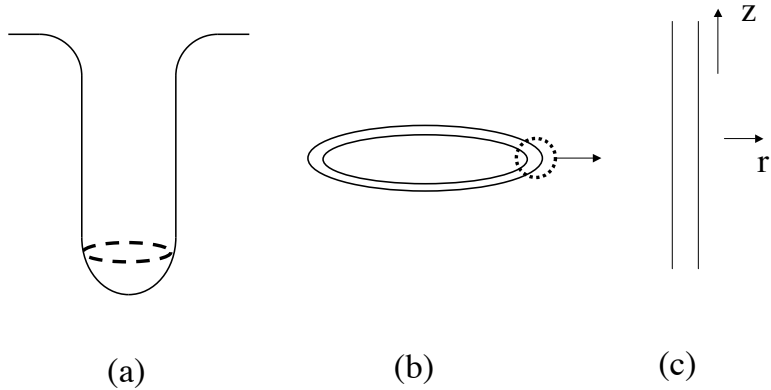}
\end{center}
\vskip -1.5truecm
\caption[x] {{\small  (a) The D1-D5 geometry for large values of $R_y$, the radius of  $S^1$; there is  
a large AdS region  (b)
The geometry for small $R_y$; the metric is close to flat outside a thin ring  
(c) In the near ring limit we approximate the segment of the ring by a  
straight line along $z$.}}
\label{fig1}
\end{figure}

\section{The CFT state}

It is important for us that the state we construct in the gravity  
description be known to be a BPS state in the dual CFT. In this section  
we review the discussion of \cite{mss} where this state was described.

\subsection{The 2-charge geometry}

We start with the 2-charge D1-D5 system. We compactify IIB string  
theory as $M_{9,1}\r M_{4,1}\times T^4\times S^1$. We wrap $n_5$ D5  
branes on $T^4\times S^1$, and $n_1$ D1 branes on $S^1$. The system has  
a large class of BPS bound states, out of which we choose a simple one  
that was first noted in \cite{bal,mm}. If we reduce the metric on $T^4$  
we find that the 6-d string metric is the same as the Einstein metric,  
so we will just call it the `metric' below.  The masses of the D1 and  
D5 branes are described by parameters $\bar Q_1,\bar Q_5$ which we will  
set to be equal
\be
\bar Q_1=\bar Q_5=\bar Q
\label{equal}
\ee
This will simplify our computations, but we expect that the state we  
construct will exist for all $\bar Q_1,\bar Q_5$.
With the choice (\ref{equal}) the dilaton is constant, and the volume  
of the $T^4$ is also constant. The metric and gauge field for the  
solution are given by
\bea
ds^2&=&-H^{-1}(dt^2-dy^2)+H f \Bigl({d\rb^2\over  
\rb^2+a^2}+d\thetab^2\Bigr)-2 {a \Qb\over H f}
(\cos^2\thetab dy d\psib +\sin^2\thetab dt d\phib)\nonumber\\
&&+H\Bigl(\rb^2+{a^2 \Qb^2 \cos^2\thetab\over H^2  
f^2}\Bigr)\cos^2\thetab d\psib^2+
H\Bigl(\rb^2+a^2-{a^2 \Qb^2 \sin^2\thetab\over H^2  
f^2}\Bigr)\sin^2\thetab d\phib^2
\label{mm}
\eea
\bea
C^{(2)}&=&-{\Qb\over H f} dt\wedge dy - {\Qb \cos^2\thetab\over H  
f}(\rb^2+a^2+\Qb) d\psib\wedge
d\phib\nonumber\\
&&-{\Qb a \cos^2\thetab\over H f} dt\wedge d\psib -{\Qb a  
\sin^2\thetab\over H f} dy\wedge d\phib
\label{mmrr}
\eea
where
\be
f=\rb^2+a^2\cos^2\thetab\,,\quad H=1+{\Qb\over f}
\ee
Here
\be
a={\bar Q\over R_y}
\ee
where $R_y$ is the radius of the $S^1$.

\subsection{The `inner' region}

  Suppose we take
\be
\epsilon\equiv {a\over \Qb^{1/2}}={\Qb^{1/2}\over R_y} <<1
\label{epsilon}
\ee
(This can be achieved by taking $R_y$ large holding all other moduli  
and the charges fixed.) We can then look at the `inner region' of this  
geometry
\be
\rb <<\sqrt{\Qb}
\label{eightt}
\ee
The metric here is
\bea
ds^2&=&-{(\rb^2+a^2\cos^2\thetab)\over
\bar Q}(dt^2-dy^2)+\bar Q(d\thetab^2+{d\rb^2\over \rb^2+a^2})\nonumber\\
&&~~~
-2a(\cos^2\thetab dy d\psib+\sin^2\thetab dt d\phib)+{\bar  
Q}(\cos^2\thetab
d\psib^2+\sin^2\thetab d\phib^2)
\label{innerr}
\eea
The change of coordinates
\be
\psi_{NS}=\psib-{a\over {\bar Q}}y, ~~~\phi_{NS}=\phib-{a\over {\bar  
Q}}t
\label{spectral}
\ee
brings (\ref{innerr}) to the form $AdS_3\times S^3$
\be
ds^2=-{(\rb^2+a^2)\over {\bar Q}}dt^2+{\rb^2\over {\bar Q}}dy^2+{\bar  
Q}{d\rb^2\over
\rb^2+a^2}+{\bar Q}(d\thetab^2+\cos^2\thetab d\psi_{NS}^2+\sin^2\thetab
d\phi_{NS}^2)
\label{innerns}
\ee
This AdS geometry is dual to a 1+1 dimensional CFT. For this CFT we can  
construct chiral primaries, which are described in the gravity picture  
by certain BPS configurations. The CFT dual to the geometry  
(\ref{innerns}) is in the Neveu-Schwarz (NS) sector. In the original  
form (\ref{innerr}) the geometry described the CFT in the Ramond (R)  
sector, and the coordinate change (\ref{spectral}) gives the gravity  
description of the `spectral flow' between the NS and R sectors  
\cite{bal,mm}.

The simplest chiral primaries can be obtained by finding normalizable  
solutions of the  supergravity equations describing linear  
perturbations around $AdS_3\times S^3$. The supergravity fields in the  
6-d theory separate into different sets (with no coupling at the linear  
level between sets). One set described an antisymmetric 2-form
$B^{(2)}_{MN}$ and a scalar $w$. We write
\be
H^{(3)}_{MNP}=\p_M  C^{(2)}_{NP}+\p_N
C^{(2)}_{PM}+\p_P C^{(2)}_{MN}, ~~~  F^{(3)}_{MNP}=\p_M  
B^{(2)}_{NP}+\p_N
B^{(2)}_{PM}+\p_P B^{(2)}_{MN}
\ee
Then the field equations for the perturbation $(B^{(2)}_{MN},w)$ are
\bea
&&F^{(3)}+\star_6 F^{(3)}+w\,H^{(3)}=0\nonumber\\
&&\Delta_6 w -{1\over 3} H^{(3)\,MNP}F^{(3)}_{MNP}=0
\label{eqtot}
\eea
Here the star operation $\star_6$,
the laplacian $\Delta_6$ and
index contractions in (\ref{eqtot}) are defined with respect to the  
metric (\ref{mm}).

\subsection{Constructing the chiral primary}

We can solve the equations (\ref{eqtot}) in the `inner region' geometry  
(\ref{innerns}) and
obtain normalizable solutions. The solution giving a chiral primary is  
\cite{exclusion,mss} (the coordinates on $S^3$ are $a,b,\dots$ and on  
the $AdS_3$ are $\mu,\nu,\dots$)
\be
w_{inner}={{e^{\displaystyle{-2i{a\over {\bar Q}}l t}}\over {\bar Q}  
(\rb^2+a^2)^l}}{\hat Y}_{NS}^{(l)}
\label{ponepre}
\ee
\be
B^{(2)}_{ab}=B\epsilon_{abc}\,\partial^c{\hat Y}^{(l)}_{NS},
~~~B^{(2)}_{\mu\nu}=[{1\over
\sqrt{{\bar Q}}}\epsilon_{\mu\nu\lambda}\p^\lambda B]~{\hat  
Y}^{(l)}_{NS}
\label{pone}
\ee
where
\be
{\hat Y}^{(l)}_{NS}=(Y^{(l,l)}_{(l,l)})_{NS}= \sqrt{ \frac{2l+1}{2
} }\frac{
e^{-2il\phi_{NS}}}{\pi} \sin^{2l}\thetab, ~~~B={1\over  
4l}{e^{\displaystyle{-2i {a\over {\bar Q}}l t }}\over (\rb^2+a^2)^l}
\label{ponepost}
\ee
In (\ref{pone}) the tensors $\epsilon_{abc}, g^{ab}$
etc are defined using the metric
on an $S^3$ with {\it unit} radius. The spherical harmonics $Y$ are  
representations of the rotation group $SO(4)\approx SU(2)\times SU(2)$  
of the sphere $S^3$, and $Y^{(l,l')}_{(m,m')}$ has quantum numbers  
$(l,m)$ in the first $SU(2)$ and $(l',m')$ in the second $SU(2)$. These  
two $SU(2)$ groups become the R symmetries of the left and right movers  
respectively in the dual CFT. The perturbation  
(\ref{ponepre}-\ref{ponepost})
gives a state in the CFT with R charges and dimensions given by
\be
j_{NS}=l, ~~h_{NS}=l, ~~~{\bar j}_{NS}=l,  ~~{\bar h}_{NS}=l
\ee
(Unbarred and barred quantities denote left and right movers  
respectively.) The quantities $j_{NS}, {\bar j}_{NS}$ are the values of  
the azimuthal quantum numbers in the two $SU(2)$ groups. The
subscript $NS$ denotes that we are in the Neveu-Schwarz sector of the  
CFT.
If we spectral flow this perturbation to the Ramond sector then we will  
get a perturbation with
\be
h=\bar h=0
\ee
which is expected, since a chiral primary of the NS sector maps under
spectral flow to a ground state of the R
sector.\footnote{The full spectral flow relations are
$h=h_{NS}-j_{NS}+{c\over 24}, ~j=j_{NS}-{c\over 12}$. Spectral flow of
the
background $|0\rangle_{NS}$ gives $h^0=h^0_{NS}-{c\over 24},
~j^0=j^0_{NS}-{c\over 12}$, so for
the
perturbation the spectral flow relations are just
$h=h_{NS}-j_{NS}, j=j_{NS}$.}

Let the CFT state dual to the perturbation
(\ref{ponepre})-(\ref{ponepost}) be called $|\psi\rangle_{NS}$, and
let $|\psi\rangle_R$ be its image under spectral flow to the Ramond
sector.

\subsection{The state  
$J^-_0|\psi\rangle_{NS}~\leftrightarrow~J^-_{- 
1}|\psi\rangle_R$}\label{thes}

  Consider again the inner region in the NS sector
coordinates (\ref{innerns}). We now wish to make the perturbation dual
to the NS sector state
\be
J^-_0|\psi\rangle_{NS}
\ee
Since the operator $J^-_0$ in the NS sector is represented by just a
simple rotation of the $S^3$, we can immediately write down the bulk
wavefunction dual to the above CFT state
\be
w_{inner}={e^{\displaystyle{-2i{a\over {\bar Q}}l t}}\over {\bar Q}  
(\rb^2+a^2)^l}Y^{(l)}_{NS}
\label{solone}
\ee
\be
B^{(2)}_{ab}=B\epsilon_{abc}\,\p^c Y^{(l)}_{NS},
~~~B^{(2)}_{\mu\nu}=[{1\over
\sqrt{{\bar Q}}}\epsilon_{\mu\nu\lambda}\partial^\lambda B]~  
Y^{(l)}_{NS}
\label{soltwo}
\ee
\be
Y^{(l)}_{NS}=(Y^{(l,l)}_{(l-1,l)})_{NS}= -\frac{\sqrt{l(2l+1)}}{\pi}
\sin^{2l-1}\thetab \cos\thetab
e^{i(-2l+1)\phi_{NS}+i\psi_{NS}},
~~~B={1\over 4l}{e^{\displaystyle{-2i {a\over {\bar Q}}l t }}\over
(\rb^2+a^2)^l}
\label{solthree}
\ee
This perturbation has
\be
j_{NS}=l-1, ~~\bar j_{NS}=l, ~~~h_{NS}=l, ~~\bar h_{NS}=l
\label{nonmatch}
\ee
The spectral flow to the R sector coordinates should give
\be
h=h_{NS}-j_{NS}=1, ~~~\bar h=\bar h_{NS}-\bar j_{NS}=0
\ee
  This spectral flowed state can be written as
\be
|\psi\rangle=J^-_{-1}|\psi\rangle_R
\ee
This is a state with nonzero $L_0-\bar L_0$, which means that
it is a state carrying  momentum P along $S^1$. It is a state in the R sector, which 
is the sector which we obtain for the CFT if we wrap D1,D5 branes around the compact directions in
our original spacetime. 

So far we have found the relevant fields only in the `inner' region  
(\ref{innerr}). But  in \cite{mss}
the perturbation equations were also solved in the `outer' region $a\ll  
r<\infty$ and it was shown that the
inner and outer region solutions agreed with each other to several  
orders in the small parameter $\epsilon$ given in (\ref{epsilon}).
This agreement was very nontrivial, and indicated that there was an  
exact solution to the perturbation problem that was smooth in the inner  
region and normalizable at spatial infinity. This exact solution would  
be a state carrying three charges: D1, D5, and one unit of momentum P.  
Since it is regular everywhere,  we learn that it is possible for   
3-charge microstate to be nonsingular and horizon free, just like  
2-charge microstates.

Even though the solution obtained in \cite{mss} was only found by  
matching inner and outer region solutions to some order in $\epsilon$,  
it was possible to guess, from the results, a closed form  for the scalar  
$w$ which would conceivably hold for all orders in $\epsilon$:
\be
w_{full}=e^{\displaystyle{-i{a\over \Qb}(t+y)}}\,e^{-i(2  
l-1)\phib}\,e^{i\psib}\,
{\sin^{2l-1}\thetab\,\cos\thetab\over (\rb^2+a^2)^l\,(\Qb+f)}
\label{fullw}
\ee

We will see that this conjecture for $w$ will help us in obtaining the  
perturbation for the 3-charge ring.

\section{The near ring limit}

In the above section we took the limit (\ref{epsilon}) which sets  
$R_y>>\sqrt{\bar Q}$; this gives the geometry of Fig.1(a) which has a  
large AdS type region. Now we will take the opposite limit
\be
R_y\ll \sqrt{\bar Q}
\label{limitp}
\ee
In this limit we get a geometry like that in Fig.1(b); we have flat  
space everywhere except around  a thin `ring'. This ring has  radius
$a=\bar Q/R_y$. Note that we have a large family of bound state D1-D5  
geometries; these arise by duality from different vibration profiles of  
the NS1-P system \cite{lm5}.  In the limit (\ref{limitp}) all these become thin  
tubes around the curves generated by the NS1-P profile. The near ring  
limit of any of these curves looks the same; thus for a local analysis  
of the perturbation equations we may start with any ring, and we have  
chosen to start with the `round' ring because it is the simplest.

In the near ring limit the following coordinates are the natural ones:  
we take a coordinate $z$ to measure length along the ring, and we  
introduce spherical polar coordinates $r,\theta,\phi$ in the  
3-dimensional space transverse to the ring. (We leave unchanged the  
coordinate along the compact directions $S^1, T^4$.)
The coordinate change from $(\rb,\thetab,\psib,\phib)$
to  $(r,\theta,\phi,z)$ is described in the Appendix. The result is
\bea
&&\rb^2={a^2 r (1-\cos\theta)\over a+r\cos\theta}\,,\quad \sin^2\thetab  
= {a-r\over a+r\cos\theta}
\,,\quad \psib=\phi \,,\quad \phib={z\over a}
\label{nrchange}
\eea
The only length scale of the near ring geometry is the parameter  
characterizing the charge density along the ring
\be
Q= {\Qb\over 2a}
\ee
The $y$ radius is given in terms of $Q$ by
\be
R_y = 2 Q
\label{ry}
\ee
The near ring region is described by
\be
r\ll a
\label{nrlimit}
\ee
 From (\ref{nrchange}) one finds, in this limit
\bea
&&\!\!\!\!\!\!\!\!\!
\rb^2\approx a r (1-\cos\theta)\,,\quad \sin^2\thetab \approx 1-  
{r\over a}(1+\cos\theta)\,,\quad f\approx 2 a r \nonumber\\
&&\!\!\!\!\!\!\!\!\! {d\rb^2\over \rb^2+a^2}+d\thetab^2={1\over 2 r  
(a+r\cos\theta)}\,\Bigl[{a^2\over a^2-r^2}dr^2 + r^2  
d\theta^2\Bigr]\approx {dr^2+r^2 d\theta^2\over 2 a r}
\eea
In this limit the  metric and RR field become
\bea
ds^2&=&-H^{-1}\,\Bigl(dt+{Q\over  
r}dz\Bigr)^2+H\,dz^2+ds^2_{TN}\nonumber\\
ds^2_{TN}&=&H^{-1}\,(dy- 
Q(1+\cos\theta)d\phi)^2+H\,(dr^2+r^2d\theta^2+r^2\sin^2\theta d\phi^2)  
\nonumber\\
C^{(2)}&=&H^{-1}{Q\over r}\,dy\wedge (dt-dz)+H^{-1}  
Q(1+\cos\theta)\,d\phi\wedge (dt-dz)
\label{nr}
\eea
where
\be
H=1+{Q\over r}
\ee

\subsection{The Taub-NUT space}

The part of the metric denoted as $ds^2_{TN}$ is Taub-NUT (TN) space:  
it is smooth due to the relation
(\ref{ry}). The TN gauge field
\be
A=-Q(1+\cos\theta) d\phi
\ee
satisfies
\be
dA = Q \sin\theta\,d\theta \wedge d\phi\,,\quad \star_3 d A = - d H
\ee
where $\star_3$ is the Hodge dual with respect to the flat  
$\mathbb{R}^3$
spanned by $r$, $\theta$, $\phi$. A convenient basis of 1-forms on TN  
is given by
$\sigmah$, $dr$, $d\theta $, $d\phi$, with
\be
\sigmah = d\yh-{1+\cos\theta\over 2}d\phi\,,\quad \yh = {y\over 2Q}
\ee
In terms of these forms the RR field strength can be written as
\be
H^{(3)}=\Omega^{(2)}\wedge (dt-dz)
\ee
with
\be
\Omega^{(2)}=-{2Q^2\over H^2 r^2}\,
\Bigl[dr\wedge \sigmah + {H r^2\over Q}\,d\sigmah\Bigl] = -2 Q\, d  
(H^{-1}\sigmah)
\label{omega2}
\ee
We choose the orientations of the 6D and TN spaces so that
\be
\epsilon_{ty\rb \thetab \psib \phib}=\epsilon_{t z y r \theta  
\phi}=1\,,\quad
\epsilon_{r\theta\phi y}=1
\ee
(and thus $\epsilon_{tz}=-1$). Then  $H^{(3)}$ is self-dual with  
respect to the 6-D metric
\be
\star_6 H^{(3)}=  H^{(3)}
\ee
and $\Omega^{(2)}$ is self-dual with respect to the 4-dimensional TN  
metric
\be
\star \Omega^{(2)}= \Omega^{(2)}
\ee
$\Omega^{(2)}$ is the unique closed and self-dual 2-form on TN.

\subsection{The scalar $w$ in the near-ring limit}

We had noted in section (\ref{thes}) that the computations of  
\cite{mss} had suggested an exact form for $w$, given in (\ref{fullw}).  

We would like to take the near ring limit of (\ref{fullw}). Remember  
that in the geometry
(\ref{mm}) the ring is spanned by the coordinate $\phib$ and its length  
is $2\pi a$. From the $\bar\phi$ dependence in (\ref{fullw}) we see  
that the
the wavelength of the perturbation $w_{full}$ in the direction of  the  
ring is
\be
\lambda={2\pi a\over 2l-1}\equiv {2\pi\over k}
\label{lambda}
\ee
We will be interested in the regime in which this wavelength is much  
shorter than the ring:
\be
  \lambda \ll a
\label{lambdasmall}
\ee
This will enable us to take our limit in such a way that we see  
oscillations of the wavefunction along the $z$ direction even when we  
take a near-ring limit and see only a short segment of the ring.  
Eqs.(\ref{lambda})
and (\ref{lambdasmall}) imply
\be
l\gg 1
\label{llarge}
\ee
We can thus approximate $2l-1\approx 2l$ in the following. By applying  
the change of coordinates
(\ref{nrchange}) and taking the limits (\ref{nrlimit}) and  
(\ref{llarge}), we find
\be
\cos\thetab=\sqrt{r(1+\cos\theta)\over a+ r\cos\theta}\approx  
\sqrt{2\over a}\,\sqrt{r}\,\cos{\theta\over 2}
\ee
and
\bea
&&{\sin^{2l-1}\thetab\over (\rb^2+a^2)^l}\approx  
\Bigl({\sin^2\thetab\over \rb^2+a^2}\Bigr)^{l}
=a^{-2l}\,\Bigl({a-r\over a+r}\Bigr)^{l}\nonumber\\
&&\approx a^{-2l}\,\Bigl(1-2{r\over a}\Bigr)^{l}\approx
  a^{-2l}\,\Bigl(1-{k\,r\over l}\Bigr)^{l}\approx a^{-2l}\,e^{-k r}
\eea
where we have used (\ref{lambda}) and the identity
$(1+\epsilon\alpha)^{1/\epsilon}\approx  e^\alpha$.
Up to an overall normalization, the near ring limit of $w_{full}$ is  
then
\be
w=e^{-i(pt + kz)}\,e^{-i(\yh-\phi)}\,\cos{\theta\over  
2}\,e^{-kr}\,{\sqrt{r}\over Q+r}
\label{nrw}
\ee
where
\be
p={a\over \Qb}={1\over 2Q}
\ee

\section{The perturbation equations}

The perturbation we seek  carries one unit of momentum along $y$ and is  
BPS: this fixes the $t$ and
$y$ dependence to be of the form $e^{-ip (t+y)}$. We also allow for a  
generic wave number $k$ along the ring direction $z$; sometimes we will  
find it convenient to write this wave number
as $k=\k/( 2 Q)$. The perturbation fields then have the form
\be
B^{(2)}_{MN}=e^{-ip(t+y)- i k z}\,{\tilde  
B}^{(2)}_{MN}(r,\theta,\phi)\,,\quad
w=e^{-ip(t+y)-i k z}\,{\tilde w}(r,\theta,\phi)
\label{genpert}
\ee

\subsection{Reducing to equations on TN}

In this subsection we reduce the equations (\ref{eqtot}) into a system  
of equations for a set
of p-forms on TN. Indices on TN are denoted by $i,j,\ldots$. Here and  
in the following $d$, $\Delta$
and $\star$ are the differential, scalar laplacian and Hodge dual on TN.
The 2-form $B^{(2)}$ reduces to a 2-form
on TN denoted by $B$, two 1-forms $a$ and $b$ and a scalar $\Phi$:
\be
B^{(2)}_{ij}=B_{ij}\,,\quad B^{(2)}_{it}=a_i\,,\quad  
B^{(2)}_{iz}=b_i\,,\quad B^{(2)}_{tz}=\Phi
\ee
Let $f^{(a)}$ and $f^{(b)}$ be the  field strengths of $a$ and $b$:
\be
f^{(a)}_{ij}=\partial_i a_j - \partial_j a_i\,,\quad  
f^{(b)}_{ij}=\partial_i b_j - \partial_j b_i
\ee
One has the identities
\be
F^{(3)}_{ijt}=f^{(a)}_{ij}-ip B_{ij}\,,\quad  
F^{(3)}_{ijz}=f^{(b)}_{ij}-i k B_{ij}\,,\quad
F^{(3)}_{itz}=\partial_i\Phi-i k a_i + ip b_i
\label{ftz}
\ee
We will need the following relations
\bea
&&g^{tt}=-H+{Q^2\over H r^2}\,,\quad g^{tz}=-{Q\over Hr}\,,\quad  
g^{zz}={1\over H}\nonumber\\
&&g^{zz}-g^{tz}=1\,,\quad g^{tt}-g^{tz}=-1
\label{gtz}
\eea
By virtue of these relations we can rewrite the 6D laplacian, acting on  
$w$, as
\be
\Delta_6 w= \Delta w -(p^2\, g^{tt} + k^2\,g^{zz}+2 p k \,g^{tz}) w  
=\Delta w- (p+k)
(p g^{tt}+ k g^{zz}) w
\label{d6}
\ee
We  also find
\be
H^{(3)\,ijt}=H^{(3)\,ijz}= -\Omega^{(2)\,i j}
\label{htz}
\ee

Using (\ref{ftz}), (\ref{d6}) and (\ref{htz}), it is easy to see that  
the equations (\ref{eqtot})
reduce to the following system of equations:
\bea
&&\label{1}\Delta w -(p+k)(p g^{tt}+ k g^{zz}) w + \Omega^{(2)\,i  
j}(f^{(a)}_{ij}+f^{(b)}_{ij}-i(p+k)
B_{ij})=0\\
&&\label{2} f^{(a)}-i p B-g^{zz}\star (f^{(b)}-i k B) -g^{tz}\star  
(f^{(a)}-i p B)+ w \Omega^{(2)}=0\\
&&\label{3}f^{(b)}-i k B +g^{tt}\star(f^{(a)}-i p B)+g^{tz}\star  
(f^{(b)}-i k B)-w \Omega^{(2)}=0\\
&&\label{4}d\Phi-i k a +ip b -\star dB=0
\eea

If we take the sum of eq. (\ref{2}) and eq. (\ref{3}), use (\ref{gtz}),  
and define
\be
K=f^{(a)}+f^{(b)}-i(p+k) B
\label{k}
\ee
we find
\be
K=\star K
\label{stark}
\ee
i.e. $K$ is a self-dual 2-form on TN. Applying $d$ to eq. (\ref{4})  
leads to
\be
p f^{(b)}- k f^{(a)}=-i d\star d B = {d\star d K\over p+k}
\label{f2f1}
\ee
Taking $p$ times eq.(\ref{3}) minus $k$ times eq.(\ref{2}), and using  
again (\ref{gtz}), gives
\be
p f^{(b)}-k f^{(a)}+\star (p f^{(b)}-k f^{(a)}) + (p g^{tt}+k  
g^{zz})\,K -(p+k)w\,\Omega^{(2)}=0
\ee
and thus, by virtue of (\ref{f2f1}) and (\ref{stark}),
\be
\Delta K + (p+k) (p g^{tt}+k g^{zz})\,K -(p+k)^2 w\,\Omega^{(2)}=0
\ee
where
\be
\Delta K = d\star d \star K + \star d \star d K =  d\star d K + \star d  
\star d K
\ee
is the TN laplacian acting on the 2-form $K$. Eq.(\ref{1}) can also be  
rewritten in form language
via the identity
\be
\Omega^{(2)\,i j} K_{ij} = 2 \star (\star \Omega^{(2)} \wedge K) = 2  
\star (\Omega^{(2)} \wedge K)
\ee

\subsection{The equations to be solved}

With all this, we have reduced the system (\ref{1}-\ref{4}) to a  
coupled system of equations
for a self-dual 2-form $K$ and scalar $w$:
\bea
&&\label{weq} \Delta w -(p+k)(p g^{tt}+ k g^{zz}) w + 2 \star  
(\Omega^{(2)} \wedge K)=0\\
&&\label{keq} \Delta K + (p+k) (p g^{tt}+k g^{zz})\,K -(p+k)^2  
w\,\Omega^{(2)}=0
\eea

Moreover, the definition of $K$ (\ref{k}) and eq.(\ref{4}) imply the  
relations
\bea
K=f^{(a)}+f^{(b)}-i(p+k) B\,,\quad {i\over (p+k)} \star d K=d\Phi-i k a  
+ip b
\label{abBPhi}
\eea
If $K$ is known, these relations determine $B$, $a$, $b$ and $\Phi$, up  
to gauge transformations.

\section{Harmonics on Taub-NUT}

We would like to solve the above equations by expanding functions on  
the Taub-NUT space in harmonics on the
$(\theta, \phi,\hat y)$ space. At the core of the Taub-NUT (i.e. at  
$r\approx 0$) this angular space has the geometry of a `round' $S^3$,  
but for larger $r$ we get a `squashed sphere'. Forms on the squashed  
sphere can be expanded in `monopole spherical harmonics', which have  
been widely studied; see for example \cite{monopole}. We will however  
find it more convenient to develop this expansion in our own notation,  
in a way that relates it closely to the expansion used for the round  
sphere in \cite{mss}.

\subsection{Symmetries of Taub-NUT}

Let us start with the metric on the round sphere
\be
ds_{S^3}^2=d\tilde \theta^2 +\cos^2\t\theta d\t\psi^2+\sin^2\t\theta  
d\t\phi^2
\label{round}
\ee
The symmetry group is $SO(4)\approx SU(2)\times SU(2)$. We write the  
elements of $SU(2)$ as $e^{\alpha_a J_a}$, with the antihermitian  
generators $J_a$ satisfying $[J_a,J_b]=-\epsilon_{abc} J_c$. Writing  
$J_{\pm}=J_1\pm i J_2$, we get
$[J_3,J_{\pm}]=\pm i J_\pm, ~[J_+,J_-]=2iJ_3$. For the first $SU(2)$  
the generators are
\bea
J_+&=&{1\over 2} e^{-i(\t\psi+\t\phi)}[\p_{\t\theta}-i\cot\t\theta  
\p_{\t\phi}+i\tan\t\theta \p_{\t\psi}]\nn
J_-&=&{1\over 2} e^{i(\t\psi+\t\phi)}[\p_{\t\theta}+i\cot\t\theta  
\p_{\t\phi}-i\tan\t\theta \p_{\t\psi}]\nn
J_3&=&-{1\over 2}[\p_{\t\psi}+\p_{\t\phi}]
\label{left}
\eea
and for the second $SU(2)$ they are
\bea
\bar J_+&=&{1\over 2} e^{i(\t\psi-\t\phi)}[\p_{\t\theta}-i\cot\t\theta  
\p_{\t\phi}-i\tan\t\theta \p_{\t\psi}]\nn
\bar J_-&=&{1\over 2} e^{-i(\t\psi-\t\phi)}[\p_{\t\theta}+i\cot\t\theta  
\p_{\t\phi}+i\tan\t\theta \p_{\t\psi}]\nn
\bar J_3&=&{1\over 2}[\p_{\t\psi}-\p_{\t\phi}]
\label{right}
\eea

To relate these generators to Taub-NUT we write the metric  
(\ref{round}) for the round $S^3$ in different coordinates. Thus define
\be
\theta=2\t\theta, ~~~\hat y\equiv\t\phi, ~~\phi=\t\phi-\t\psi
\label{map}
\ee
This gives
\be
ds^2_{S^3}={1\over 4}d\theta^2+{1\over 4}\sin^2\theta d\phi^2+
\Bigl[d\hat y-{1\over 2}(1+\cos\theta) d\phi\Bigr]^2
\ee
The generators (\ref{left}),(\ref{right}) become
\bea
J_+&=&{1\over 2}e^{-i(2\hat y-\phi)}[2\p_\theta-i(\cot{\theta\over  
2}+\tan{\theta\over 2})\p_\phi-i\cot{\theta\over 2} \p_{\hat y}]\nn
J_-&=&{1\over 2}e^{i(2\hat y-\phi)}[2\p_\theta+i(\cot{\theta\over  
2}+\tan{\theta\over 2})\p_\phi+i\cot{\theta\over 2} \p_{\hat y}]\nn
J_3&=&-{1\over 2}\p_{\hat y}
\label{leftp}
\eea
\bea
\bar J_+&=&{1\over 2}e^{-i\phi}[2\p_\theta-i(\cot{\theta\over  
2}-\tan{\theta\over 2})\p_\phi-i\cot{\theta\over 2} \p_{\hat y}]\nn
\bar J_-&=&{1\over 2}e^{i\phi}[2\p_\theta+i(\cot{\theta\over  
2}-\tan{\theta\over 2})\p_\phi+i\cot{\theta\over 2} \p_{\hat y}]\nn
\bar J_3&=&-{1\over 2}[\p_{\hat y}+2\p_{\phi}]
\label{rightp}
\eea

In the Taub-NUT metric if we fix $r$ then we get a 3-dimensional  
surface with metric of the form
\be
ds^2=A(d\theta^2+\sin^2\theta d\phi^2)+4B\Bigl[d\hat y-{1\over  
2}(1+\cos\theta)d\phi\Bigr]^2
\label{mgeneral}
\ee
At the center of Taub-NUT we get $A=B$, and the metric becomes that of  
a round $S^3$. For larger $r$ we have $A\ne B$, and this gives the  
squashed sphere. We can now check that the vector fields (\ref{rightp})  
are Killing vectors of (\ref{mgeneral}),
for all $A,B$. But out of the vector fields (\ref{leftp}) only $J_3$ is  
a Killing vector if $A\ne B$. Thus the $SU(2)\times SU(2)$ symmetry of  
the round sphere is broken to $U(1)\times SU(2)$.

\subsection{Harmonics on the squashed sphere}

On the round sphere we can expand any form in spherical harmonics,  
which are characterized by quantum numbers
$(j,m), (j',m')$ in the two $SU(2)$ factors. On the squashed sphere, we  
can use the {\it same} functions, in the following sense.
We take the map from the squashed $S^3$ to the round $S^3$ which sends  
each point $(\theta, \phi,\hat y)$ on the former to the point with the  
same coordinates on the latter. The harmonics on the round $S^3$ then  
give harmonics on the squashed $S^3$ via the pullback under this map.  
These pulled back harmonics can be used to expand any form on the  
squashed sphere, though the harmonics  are not orthogonal to each other  
as they were on the round sphere.

The quantum numbers $m$ and $(j',m')$ correspond to symmetries of the  
squashed sphere, and so are `good' quantum numbers in the sense that we  
can restrict all terms in an equation to have the same values of these  
numbers. On the other hand a form of  order  $p$ on the round $S^3$ was  
characterized by {\it four} quantum numbers, $(j,m),(j',m')$. In the  
latter case the quantum numbers uniquely specify the form. A form on  
the squashed sphere will therefore be a sum
\be
\omega_{(m,j',m')}=\sum_j C_j \, \omega_{(j,m),(j',m')}
\label{pullback}
\ee
It turns out however that if $\omega$ is a p-form then for its  
harmonics on the round sphere we must have $|j-j'|\le p$. This tells us  
that the sum in (\ref{pullback}) will be a finite one, and this makes  
the expansion in harmonics useful for the squashed sphere.

As an application of this approach consider the scalar $w$ in  
(\ref{nrw}): its angular dependence is captured by the function
\be
\omega_0=e^{-i(\yh-\phi)}\,\cos{\theta\over 2}=e^{-i \psit}\,\cos\thetat
\label{o0}
\ee
All scalars on $S^3$ have quantum numbers $(j,m),(j,m')$; i.e. $j=j'$.  
 From the $\psit$ dependence of (\ref{o0}) we find that
$m={1\over 2}, m'=-{1\over 2}$. The lowest $j$ this can come from is  
$j={1\over 2}$, so we look at the scalar spherical harmonic on the  
round $S^3$ given by the quantum numbers
$({1\over 2}, {1\over 2}), ({1\over 2}, -{1\over 2})$. Such harmonics  
were given explicitly in \cite{mss}, and we find that indeed the  
function (\ref{o0}) is proportional to the required scalar harmonic.

Now consider 1-forms. On the round $S^3$, there are two kinds of  
1-forms. The first kind are obtained by just applying $d$ to the scalar  
harmonics, so these have quantum numbers $(j,m),(j,m')$. The second  
kind have $j-j'=\pm 1$, so they come in two varieties: with quantum  
numbers
$(j+1,m),(j,m')$, and $(j,m),(j+1,m')$. Let us examine these 1-forms  
for our problem.

Since $m, j', m'$ are good quantum numbers for the problem these must  
be the same for the 1-forms as for the scalar $w$. Thus the first kind  
of 1-form must be
\be
d \omega_0=-e^{-i(\hat y-\phi)}\Bigl[
{1\over 2}\sin{\theta\over 2} d\theta +i \cos{\theta\over 2}(d\hat  
y-d\phi)\Bigr]
\label{dzero}
\ee
For the second kind of 1-form we find only one set of quantum numbers  
that are consistent with the given $m,j',m'$: the set
$({3\over 2}, {1\over 2}),({1\over 2}, -{1\over 2})$. The corresponding  
harmonic was constructed in \cite{mss}
\be
\tilde \omega_1 = e^{-i\psit}\,[\sin\thetat\,d\thetat - i
\cos\thetat (3\cos^2\t\theta-1) d\psit - 3 i \cos\thetat \sin^2\thetat  
d\phit]
\ee
In the coordinates $(\theta,\hat y,\phi)$ this is
\be
\t \omega_1=e^{-i(\hat y-\phi)}\Bigl[{1\over 2} \sin{\theta\over 2}  
d\theta+i\cos{\theta\over 2}(3\cos^2{\theta\over 2}-1)d\phi-2i  
\cos{\theta\over 2}d\hat y\Bigr]
\label{done}
\ee

\subsection{Decomposing along base and fiber}\label{deco}

At this stage we may think of expanding the angular components of our  
1-forms  using (\ref{dzero}) and (\ref{done}), and the $dr$ component  
using the scalar harmonic $\omega_0$. But actually we can do better, by  
exploiting the spherical symmetry of the background in the  
$r,\theta,\phi$ space. The Taub-NUT has such a spherical symmetry,  
though any  choice of coordinates prevents this symmetry from being  
manifest.

Consider the squashed sphere at any $r$. This space can be regarded as  
a $S^1$ fiber (parameterized by $\hat y$) over a $S^2$ base  
(parametrized by $\theta,\phi$). We can geometrically identify the  
fiber direction over any point, and thus also the 2-plane orthogonal to  
the fiber. We can thus decompose any 1-form into two parts:  
$\omega_1=\alpha+\beta$. The part $\alpha$ will have no component along  
the base; thus $\langle v,\alpha\rangle=0$ for all $v$ perpendicular to  
$\p_{\hat y}$.
The part $\beta$ will have no component along the fiber; thus $\langle  
\p_{\hat y},\beta\rangle=0$. We find that $\alpha$ must be proportional  
to
\be
\hat \sigma=d\hat y-{1+\cos\theta\over 2} d\phi
\ee
while $\beta$ is just characterized by having no term proportional to  
$d\hat y$.

Let us now apply this decomposition to our 1-form. The part $\alpha$  
can be written as $\alpha=f\hat\sigma$ where $f$ is a function on the  
squashed sphere. This function must carry the quantum numbers $(m',  
j,m)=({1\over 2}, {1\over 2}, -{1\over 2})$. Since $f$ is a scalar it  
must have $j=j'$ and so it must actually be proportional to the scalar  
harmonic that is the pullback of $({1\over 2}, {1\over 2}), ({1\over  
2}, -{1\over 2})$, which is just $\omega_0$. So we find that  
$\alpha=f_0(r) \omega_0 \hat\sigma$.
To find $\beta$ we take the linear combination of (\ref{dzero}),  
(\ref{done}) which has no component $d\hat y$. A conveniently  
normalized choice for this combination is
\be
\omega_1=-{4\over 3}\Bigl[d\omega_0-{1\over 2}\t  
\omega_1\Bigr]=e^{-i(\hat y-\phi)}\sin{\theta\over  
2}[d\theta-i\sin\theta d\phi]
\label{o1}
\ee

To summarize, our 1-form must have the form  
$f_0(r)\omega_0\hat\sigma+f_1(r)\omega_1$.

\subsection{Some relations on forms}

A decomposition of the type $\omega_1=\alpha+\beta$ which we did for  
1-forms can be done for any p-form $\omega$ on the squashed 3-sphere.  
The  $\yh$ dependence of our forms is $e^{-i\hat y}$. Using this fact   
we find
\be
d\omega = -i{\sigmah}\wedge \omega + D\omega
\ee
where
\be
D \omega \equiv d_2 \omega -i {1+\cos\theta\over 2}d\phi\wedge \omega
\ee
is the covariant derivative of $\omega$ and $d_2$ denotes the  
differential with respect to $\theta,\phi$. The square of $D$ is  
proportional to the monopole
field strength
\be
D^2 \omega = \Bigl (i{\sin\theta\over 2}\,d\theta\wedge  
d\phi\Bigr)\wedge\omega
\ee

If we denote by $\star_2$ the Hodge dual with respect to the $S^2$  
metric,\footnote{Note that  $(\star_2)^2= -1$ on 1-forms. We have  
$\epsilon_{\theta\phi}=1$.} the monopole harmonics $\omega_0$ and  
$\omega_1$ satisfy
\bea
&&\star_2 \omega _1 = -i \omega_1\nonumber\\
&&D\omega_0 = -{1\over 2}\,\omega_1\nonumber\\
&&D^2 \omega_0 = i\,{\sin\theta\over 2}\,\omega_0\,d\theta\wedge  
d\phi\nonumber\\
&&D\omega_1= -2 D^2 \omega_0 = -i\sin\theta\,\omega_0\,d\theta\wedge  
d\phi
\label{ooidentities}
\eea

\subsection{The 2-form $K$}

We can use the structure above to write a general ansatz for the 2-form  
$K$. Any self-dual 2-form on TN
can be written as
\be
dr\wedge \tilde \omega + \star (dr\wedge \tilde \omega)
\ee
where $\tilde \omega$ is a 1-form on TN. The form $dr$ has all angular quantum  
numbers zero, so the quantum numbers of the perturbation must be  
carried by  $\t \omega$. But we have seen in section (\ref{deco}) that any such 1-form,  
with quantum numbers
$(m', j,m)=({1\over 2}, {1\over 2}, -{1\over 2})$, is of the form
\be
\tilde \omega = f_0(r)\,\omega_0\hat\sigma+f_1(r)\,\omega_1
\label{breakup}
\ee
The 2-form $K$ is self-dual, as a form on TN, and depends on $t$ and  
$z$ as in (\ref{genpert}): it can thus
be written as
\be
K=e^{-i(t+\k z)/(2 Q)}\,(K_0+K_1)
\ee
where the $K_0, K_1$ parts correspond to the first and second parts on  
the RHS of (\ref{breakup})
\bea
K_0&=&f_0(r)\,\omega_0 [dr\wedge \sigmah + \star(dr\wedge \sigmah)]=
f_0(r)\,\omega_0\,\Bigl[dr\wedge \sigmah + {H r^2\over Q}  
\,d\sigmah\Bigr]\nonumber\\
K_1 &=& f_1(r)\,
[dr\wedge \omega_1 + \star (dr\wedge \omega_1)]=
f_1(r)\, \Bigl[dr\wedge \omega_1  -i {2Q\over H} \,\sigmah\wedge  
\omega_1 \Bigr]
\label{ansatz}
\eea

Note that we have  used only the scalar $\omega_0$ and the 1-form  
$\omega_1$ in our expansion, and avoided a separate coefficient  
function for the 1-form $d\omega_0$

\section{The radial equations and their solution}

The ansatz (\ref{ansatz}) reduces the unknowns to two functions of $r$:  
$f_0(r)$ and $f_1(r)$. In this section we will derive the system of  
differential equations these functions have to satisfy, and then see  
how they are solved.

\subsection{Obtaining the radial equations}

Let us start from eq. (\ref{weq}). We note that
\be
\Omega^{(2)}\wedge K_1 =0
\ee
and that, by comparing (\ref{ansatz}) with (\ref{omega2}),
\be
K_0=-f_0 \,\omega_0\,{(r H)^2\over 2Q^2}\,\Omega^{(2)}
\ee
Thus
\be
\star (\Omega^{(2)}\wedge K) = -e^{-i(t+\k z)/(2 Q)}\,f_0  
\,\omega_0\,{(r H)^2\over 2Q^2}\,
\star(\Omega^{(2)}\wedge \Omega^{(2)})
\label{o2k}
\ee
An easy computation gives
\be
\Omega^{(2)}\wedge \Omega^{(2)}={4 Q^3 \over H^3  
r^2}\,\sin\theta\,dr\wedge d\theta\wedge d\phi
\wedge \sigmah
\ee
and
\be
\star (\Omega^{(2)}\wedge \Omega^{(2)}) ={2 Q^2\over (H r)^4}
\ee
Using this in (\ref{o2k}) we find
\be
\star (\Omega^{(2)}\wedge K) = -e^{-i(t+\k z)/(2 Q)}\,f_0  
\,\omega_0\,{1\over (H r)^2}
\label{weqtwo}
\ee
Using the expression for $w$ in (\ref{nrw}), it is also straightforward  
to compute (for
example with the help of Mathematica)
\be
\Delta w -(p+k)(p g^{tt}+ k g^{zz}) w = -e^{-i(t+\k z)/(2 Q)}\,e^{-\k  
r/(2Q)}\,
{\sqrt{r}\over ( H r)^4}\,[(3+\k) Q + (1+\k) r]\,\omega_0
\label{weqone}
\ee
Using (\ref{weqtwo}) and (\ref{weqone}), we see that equation  
(\ref{weq}) becomes
\be
e^{-\k r/(2Q)}\,{\sqrt{r}\over ( H r)^2}\,[(3+\k) Q + (1+\k) r]+2 f_0=0
\label{weqbis}
\ee

Let us now turn to eq. (\ref{keq}). We first need to compute
\bea
\Delta K_a &=& \star d \star d K_a + d \star d K_a \nonumber\\
&=& -\Bigl(\nabla^k \nabla_k K_{a\,ij} + [\nabla^k,\nabla_i]
K_{a\,jk}
-[\nabla^k ,\nabla_j]K_{a\,i k}\Bigr)\,dx^i\wedge dx^j
\label{laplki}
\eea
for $a=0,1$. The covariant derivatives and index contractions in the  
second line of (\ref{laplki})
are done with the TN metric. A lengthy but straightforward computation,  
that makes use of identities
(\ref{ooidentities}), leads to
\bea
\!\Delta K_0 &\!\!\!=\!\!\!& -\omega_0\,[dr\wedge \sigmah + \star  
(dr\wedge \sigmah)]\Bigl[{f_0''\over H}
+2{f_0'\over H r}-{r^4+4Qr^3+16Q^2r^2-8Q^3r+3Q^4\over 4Q^2  
r(Q+r)^3}\,f_0\Bigr]\nonumber\\
&&-{i\over 2Q}\,[dr\wedge \omega_1 + \star (dr\wedge  
\omega_1)]\,{f_0\over Hr}\nonumber\\
\!\Delta K_1 &\!\!\!=\!\!\!& -[dr\wedge \omega_1 + \star (dr\wedge  
\omega_1)]\Bigl[{f_1''\over H}
+2{f_1' Q\over (H r)^2}-{r^4+4Qr^3+8Q^2r^2+16Q^3r+3Q^4\over 4Q^2  
r(Q+r)^3}\,f_1\Bigr]\nonumber\\
&&+i 4Q\,[dr\wedge \sigmah + \star (dr\wedge \sigmah)]\,{f_1\over  
(Hr)^3}
\eea
The full wave operator is
\be
\Delta_6 K_a\equiv \Delta K_a + (p+k)(p g^{tt} + k  
g^{zz})\,K_a\,,\,\,a=0,1
\ee
and we find
\bea
\Delta_6 K_0 &\!\!\!=\!\!\!&-\omega_0\,[dr\wedge \sigmah +\star  
(dr\wedge \sigmah)]\Bigl[
{f_0''\over H} +2{f_0'\over H r}-
\Bigl({r^4+4 Q r^3+16 Q^2 r^2 -8 Q^3 r + 3 Q^4 \over r(H r)^3}  
\nonumber\\
&&\qquad -(1+\k)\,{2Q+(1-\k) r\over H r}\,\Bigl)\,{f_0\over  
(2Q)^2}\Bigr]
-{i\over2Q}\,[dr\wedge \omega_1 + \star (dr\wedge \omega_1)]\,{f_0\over  
Hr}\nonumber\\
\Delta_6 K_1 &\!\!\!=\!\!\!& -[dr\wedge \omega_1 + \star (dr\wedge  
\omega_1)]\Bigl[{f_1''\over H} +2{f_1' Q\over (H r)^2}-\Bigl({r^4+4 Q  
r^3+8 Q^2 r^2 +16 Q^3 r + 3 Q^4 \over r( H r)^3}\nonumber\\
&&\qquad -(1+\k)\,{2Q+(1-\k) r\over H r}\,\Bigl)\,{f_1\over (2  
Q)^2}\Bigr]
+i 4Q\,[dr\wedge \sigmah + \star (dr\wedge \sigmah)]\,{f_1\over (Hr)^3}
\eea
In (\ref{keq}) the first two terms constitute $\Delta_6$; thus the last  
term will act as a `source term' for this Laplacian. The source term is
\be
-(p+k)^2\,w\Omega^{(2)}= {(1+\k)^2\over 2}\,e^{-i(t+\k z)/(2  
Q)}\,e^{-\k r/(2Q)}{\sqrt{r}\over (H r)^3}\,\omega_0\,[dr\wedge \sigmah  
+ \star (dr\wedge \sigmah)]
\ee
Collecting the terms proportional to $\omega_0\,[dr\wedge \sigmah +  
\star (dr\wedge \sigmah)]$ and to
$[dr\wedge \omega_1 + \star (dr\wedge \omega_1)]$ in eq. (\ref{keq}),  
we find the following system
of equations for $f_0$ and $f_1$:
\bea
&&\label{f0eq}
\!\!\!\!\!\!\!
{f_0''\over H} +2{f_0'\over H r}-\Bigl({r^4+4 Q r^3+16 Q^2 r^2 -8 Q^3 r  
+ 3 Q^4 \over r( H r)^3}
-(1+\k){2Q+(1-\k) r\over H r}\Bigl)\,{f_0\over (2 Q)^2}\nonumber\\
&&\qquad -i\,4Q\,{f_1\over (H r)^3}-{(1+\k)^2\over 2}\,e^{-\k r/(2Q)}\,
{\sqrt{r}\over (H r)^3}=0\\
&&\label{f1eq}
\!\!\!\!\!\!\!{f_1''\over H} +2{f_1' Q\over (H r)^2}-\Bigl(
{r^4+4 Q r^3+8 Q^2 r^2 +16 Q^3 r + 3 Q^4 \over r( H  
r)^3}-(1+\k){2Q+(1-\k) r\over H r}\Bigl)\,{f_1\over (2 Q)^2}\nonumber\\
&&\qquad +{i\over 2Q}\,{f_0\over (H r)}=0
\eea

\subsection{Solving the radial equations}

Eq. (\ref{weqbis}) can be readily solved for $f_0$, giving:
\be
f_0 = -e^{-kr}\,{(3+\k)Q+(1+\k)r\over 2}\,{\sqrt{r}\over (H r)^2}
\label{solf0}
\ee
By substituting $f_0$ into (\ref{f0eq}) we can derive $f_1$. With the  
help of
Mathematica we compute:
\bea
&&\!\!\!\!\!{f_0''\over H} +2{f_0'\over H r}-\Bigl({r^4+4 Q r^3+16 Q^2  
r^2 -8 Q^3 r + 3 Q^4 \over r( H r)^3}-(1+\k){2Q+(1-\k) r\over H  
r}\Bigl)\,{f_0\over (2 Q)^2}\nonumber\\
&&\qquad\qquad  =e^{-k r}\,{2+(1+\k)^2\over 2}\,{\sqrt{r}\over (H r)^3}
\eea
and thus, from (\ref{f0eq}), we obtain
\be
f_1 =-{i\over 4Q}\,e^{-k r}\,\sqrt{r}
\ee
Eq. (\ref{f1eq}) is a consistency condition for our previously  
determined values of $f_0$ and
$f_1$. By Mathematica we compute
\bea
&&\!\!\!\!\!
{f_1''\over H} +2{f_1' Q\over (H r)^2}-\Bigl(
{r^4+4 Q r^3+8 Q^2 r^2 +16 Q^3 r + 3 Q^4 \over r( H  
r)^3}-(1+\k){2Q+(1-\k) r\over H r}\Bigl)\,{f_1\over (2 Q)^2}\nonumber\\
&&\qquad\qquad  ={i\over  
4Q}\,e^{-kr}\,[(3+\k)Q+(1+\k)r]\,{\sqrt{r}\over (H r)^3}
\eea
Substituting this in (\ref{f1eq}) and using (\ref{solf0}), we see that   
(\ref{f1eq}) is satisfied.

To summarize,  we have found the solution
\bea
\label{w}w&\!\!\!=\!\!\!&e^{-i (p t +k z)}\,e^{-kr}\,{\sqrt{r}\over  
Q+r}\,\omega_0\\
K&=&e^{-i (p t +k z)}\,(K_0+K_1)\nonumber\\
K_0&\!\!\!=\!\!\!&-e^{-k r}\,{\sqrt{r}\over 2 (Q+r)^2}[(3Q+r)+\k  
(Q+r)]\omega_0
\Bigl[dr\wedge \sigmah + {H r^2\over 2 Q} \,\sin\theta d\theta\wedge  
d\phi\Bigr]\nonumber\\
K_1 &\!\!\!=\!\!\!& -i e^{- kr}\,{\sqrt{r}\over 4Q}
\Bigl[dr\wedge \omega_1 -i {2Q\over H} \,\sigmah\wedge \omega_1 \Bigr]
\eea

Knowing $K$ we can  derive the values of $a$, $b$, $B$ and $\Phi$.  
Before we do this we discuss the gauge
invariance of our problem.

\subsubsection{Gauge invariance}

One can easily check, using the identities (\ref{ftz}), that the  
following two gauge transformations leave all
the components of $F^{(3)}$ invariant:
\bea
B\to B + d\lambda^{(1)}\,,\quad a\to a+ip\,\lambda^{(1)}\,,\quad b\to  
b+ i k\,\lambda^{(1)}
\label{gauge1}
\eea
\bea
\!\!\!\!\!\!\!\!\Phi\to \Phi + \lambda^{(0)}\,,\quad a\to a +  
d\lambda^{(0)}_a\,,\quad b\to b + d\lambda^{(0)}_b\quad{\rm  
with}\quad\lambda^{(0)}-i k\,\lambda^{(0)}_a+ip\,\lambda^{(0)}_b=0
\label{gauge2}
\eea
Here $\lambda^{(1)}$ is a 1-form and $\lambda^{(0)}$,   
$\lambda^{(0)}_a$ and $\lambda^{(0)}_b$ are 0-forms on TN.
The 2-form $K$ is gauge invariant.

\subsubsection{Deriving the gauge fields}
By making use of the transformation (\ref{gauge2}), we can set $\Phi=0$.
Then the second equation in (\ref{abBPhi}) implies
\be
b-\k a = (2Q)^2 {\star d K\over 1+\k}
\label{abone}
\ee
One can compute
\bea
\!\!\!\!\!
{\star d K\over 1+\k} = e^{-i (p t +k z)}\,e^{- k  
r}\,\sqrt{r}\,\Bigl[-i {1\over 8 Q^2}\,\omega_1+
i {1\over 8 Q^2\,r}\,
\omega_0\, dr +{1\over 4 (Q+r)^2}\Bigl(1-\k {r\over  
Q}\Bigr)\,\omega_0\, \sigmah\Bigr] \nonumber\\
\eea
Then a solution of (\ref{abone}) for $a$ and $b$ is
\bea
a &=& e^{-i (p t +k z)}\,e^{- k r}\,\sqrt{r}\,{Q r\over  
(Q+r)^2}\,\omega_0\,\sigmah\nonumber\\
b&=& e^{-i (p t +k z)}\,e^{- k r}\,\sqrt{r}\,\Bigl[ {i\over 2 r  
}\omega_0\,dr -{i\over 2}\,\omega_1
+{Q^2\over (Q+r)^2}\, \omega_0\, \sigmah \Bigr]
\label{ab}
\eea
By picking this solution we have fixed the gauge freedom implied by the  
transformation (\ref{gauge1}).

Substituting these values of $a$ and $b$ into the first equation in  
(\ref{abBPhi}) we derive $B$
\be
B=e^{-i (p t +k z)}\,e^{- k r}\,\sqrt{r}\,\Bigl[{1\over 2} dr\wedge  
\omega_1 -{i r\over 2}\,
\omega_0 \,\sin\theta d\theta\wedge d\phi\Bigr]
\label{B}
\ee

\section{Regularity of the solution}

We show that the solution given in (\ref{w},\ref{ab},\ref{B}) is both  
regular and normalizable.

\subsection{Normalizability}\label{norm}

For $k>0$ normalizability is guaranteed by the exponential fall off  
$e^{- k r}$. Note however that
waves with $k\le 0$ give rise to non-normalizable perturbations. This  
is obvious for $k<0$. For
$k=0$ let us look, for example, at the scalar $w$: its large $r$  
behavior is
\be
w\approx e^{-i p t}\,{1\over \sqrt{r}}\,\omega_0
\ee
and thus
\be
|w|^2\sim 1/r
\ee
Since the volume element of the space transverse to the ring  grows as  
$\sim r^2dr$ for large $r$, the norm of $w$ is quadratically divergent  
at $r\to\infty$. This shows that for $k=0$ the perturbation leaks out  
to the center of the ring ($r\sim a$)
and does not stay confined to the vicinity of the tube.\footnote{If we  
construct the exact wavefunction for the ring (instead of just  
constructing it for the near ring limit) then we expect to have a  
solution normalizable at spatial infinity, since the state exists in  
the dual CFT.} For  $k>0$ the wavefunction becomes confined closer to  
the ring, and in the limit (\ref{nrlimit}) we find a normalizable  
solution in the near ring limit.

The fact that positive and negative $k$ behave differently is to be  
expected; the 2-charge background does not have the symmetry  
$z\leftrightarrow -z$. In the NS1-P picture the geometry is created by  
a  string carrying a wave, and the strands of the string carry momentum  
  along the ring,  thus breaking the $z\leftrightarrow -z$ symmetry. In  
\cite{supertubegms} it was found that there are `left-moving' non-BPS  
perturbations that move in one direction along the ring, while  
`right-moving' perturbations
create time independent distortions of the 2-charge geometry. For our  
present problem note that in $w_{full}$ (eq. (\ref{fullw})) we have  
$l\ge {1\over 2}$, so we must have $k \ge 0$, and negative values of  
$k$ do not appear.\footnote{Spherical harmonics for the scalar have  
$l=0, {1\over 2}, 1,\dots$, but for $l=0$ we get zero if we apply  
$J_0^-$, and so we cannot construct the required perturbation of  
section (\ref{thes}).} We need to take large $|k|$ to be able to use  
the `straight segment' limit of the ring so the case $k=0$ is not  
relevant for our discussion, and we naturally find ourselves at large  
positive $k$.

\subsection{Regularity at $\theta=0,\pi$}

Let us now consider the regularity of the solution.
The fields $w$, $a$, $b$ and $B$ are manifestly regular away from the  
points where our system
of coordinates degenerates. This degeneration happens  at $\theta=0$ or  
$\pi$ and at $r=0$. Around $\theta=0,\pi$ it is
convenient to change to $S^3$ coordinates (\ref{map}). From the  
expression of $\omega_0$ in
(\ref{o0}) it is apparent that $\omega_0$ is regular: indeed for  
$\thetat=\pi/2$, where the
$\psit$ coordinate degenerates, the coefficient of $e^{-i\psit}$  
vanishes. The second
identity in (\ref{ooidentities}) expresses $\omega_1$ as the covariant  
derivative of $\omega_0$, and thus $\omega_1$ is regular too. The  
1-form along the fiber $\sigmah$ can be expressed in $S^3$
coordinates as
\be
\sigmah=\sin^2\thetat\,d\phit+\cos^2\thetat\,d\psit
\ee
which is also regular. Since the angular dependence of $w$, $a$, $b$  
and $B$ is entirely
expressed in terms of $\omega_0$, $\omega_1$ and $\sigmah$, this proves  
that our solution is
regular at $\theta=0,\pi$.

\subsection{Regularity at $r=0$}
\label{regularityr=0}
At $r\to 0$ the TN space becomes flat $\mathbb{R}^4$: the change of  
coordinates that brings the TN metric into explicitly flat form is  
(\ref{map}) for the angular variables and
\be
r={\rho^2\over 4 Q}
\ee
for the radial coordinate. In these coordinates the $r\to 0$ limit of  
$w$ is
\be
w\sim \rho\,e^{-i \psit}\,\cos\thetat =  x_1-ix_2
\ee
where  $x_i$, with $i=1,\ldots,4$ are Cartesian coordinates\footnote{Explicitly,
\be
x_1+ix_2 = \rho\,e^{i\psit}\,\cos\thetat\,,\quad  
x_3+ix_4=\rho\,e^{i\phit}\,\sin\thetat
\ee} in  
$\mathbb{R}^4$.
This shows that $w$ is regular at $r\to 0$. Similarly, the gauge fields  
$a$ and $B$ behave like
\bea
a&\!\!\!\sim\!\!\!&  
\rho^3\,e^{- 
i\psit}\,\cos\thetat\,(\sin^2\thetat\,d\phit+\cos^2\thetat\,d\psit)=
(x_1-ix_2)\,[x_1 dx_2 - x_2 dx_1+x_3 dx_4 - x_4 dx_3]\nonumber\\
B&\!\!\!\sim\!\!\!&-i\rho^2\,e^{-i\psit}\,\sin\thetat\,[i d\rho\wedge  
d\thetat
+\cos\thetat(\sin\thetat d\rho + \rho \cos\thetat d\thetat)\wedge  
(d\phit-d\psit)]\nonumber\\
&\!\!\!=\!\!\!&-i[(x_1-ix_2)(dx_1 \wedge dx_2 + dx_3\wedge dx_4)-i  
\sum_i x_i dx_i \wedge (dx_1-idx_2)]
\eea
and are hence regular. Regularity of $b$ is not manifest in the form in  
which it appears in (\ref{ab}).
This form was derived after making the arbitrary gauge choice $\Phi=0$.  
By using the transformation
(\ref{gauge2}) we can change $\Phi$ and $b$ and write them in an  
explicitly smooth form; since
$a$ was already shown to be smooth, we can take $\lambda^{(0)}_a=0$ in  
(\ref{gauge2}) and leave it
unchanged. If we choose
\be
\lambda^{(0)}=-e^{-i (p t +k z)}\,e^{-k r}\,{\sqrt{r}\over  
2Q}\,\omega_0\,,\quad \lambda^{(0)}_b =
{i\over p}\,\lambda^{(0)}
\ee
in (\ref{gauge2}), then the fields $\Phi$ and $b$ are changed into
\bea
\Phi&=&-e^{-i (p t +k z)}\,e^{-k r}\,{\sqrt{r}\over  
2Q}\,\omega_0\nonumber\\
b&=& e^{-i (p t +k z)}\,e^{-k r}\,\sqrt{r}\,\Bigl[i  
k\,\omega_0\,dr+\Bigl({Q^2\over (Q+r)^2}-1\Bigr)
\,\omega_0\,\sigmah\Bigr]
\eea
At $r\to 0$ both $\Phi$ and $b$ are now explicitly regular:
\bea
\Phi&\!\!\!\sim\!\!\!& \rho\,e^{-i \psit}\,\cos\thetat =   
x_1-ix_2\nonumber\\
b&\!\!\!\sim\!\!\!& \rho\,e^{-i\psit}\,\cos\thetat\Bigl[{i k\over  
2Q}\rho\,d\rho-{\rho^2\over 2Q^2}\,
(\sin^2\thetat\,d\phit+\cos^2\thetat\,d\psit)\Bigl]\nonumber\\
&\!\!\!=\!\!\!&(x_1-ix_2)\,\Bigl[{i k\over 2Q}\sum_i x_i dx_i-{x_1 dx_2  
- x_2 dx_1+x_3 dx_4 - x_4 dx_3\over 2 Q^2}\Bigl]
\eea

\section{Summary of the solution}

We summarize here the full solution for $w$ and $B^{(2)}_{MN}$, in the  
gauge of section
\ref{regularityr=0} where all fields are regular:
\bea
w&=&e^{-{1\over 2Q}(t+y)}\, e^{i(\phi-kz)}\,\cos{\theta\over  
2}\,e^{-kr}\,{{r}^{1/2}\over Q+r}\nonumber\\
B^{(2)}_{tz}&=&- e^{-{1\over 2Q}(t+y)}\,  
e^{i(\phi-kz)}\,\cos{\theta\over 2}\,e^{-kr}\,
{r^{1/2}\over 2 Q}\nonumber\\
B^{(2)}_{yt}&=&e^{-{1\over 2Q}(t+y)}\, e^{i(\phi-kz)}\,\cos{\theta\over  
2}\,e^{-kr}\,{r^{3/2}\over 2 (Q+r)^2}\nonumber\\
B^{(2)}_{\phi t}&=&-e^{-{1\over 2Q}(t+y)}\,  
e^{i(\phi-kz)}\,\cos^3{\theta\over 2}\,e^{-kr}\,{Q\,r^{3/2}\over  
(Q+r)^2}\nonumber\\
B^{(2)}_{yz}&=&e^{-{1\over 2Q}(t+y)}\, e^{i(\phi-kz)}\,\cos{\theta\over  
2}\,e^{-kr}\,{r^{1/2}\over 2 Q}\,
\Bigl[{Q^2\over (Q+r)^2}-1\Bigr]\nonumber\\
B^{(2)}_{\phi z}&=&-e^{-{1\over 2Q}(t+y)}\,  
e^{i(\phi-kz)}\,\cos^3{\theta\over 2}\,e^{-kr}\,r^{1/2}\,
\Bigl[{Q^2\over (Q+r)^2}-1\Bigr]\nonumber\\
B^{(2)}_{rz}&=&i k\,e^{-{1\over 2Q}(t+y)}\,  
e^{i(\phi-kz)}\,\cos{\theta\over 2}\,e^{-kr}\,r^{1/2}\nonumber\\
B^{(2)}_{r\theta}&=&e^{-{1\over 2Q}(t+y)}\,  
e^{i(\phi-kz)}\,\sin{\theta\over 2}\,e^{-kr}\,{r^{1/2}\over 2}
\nonumber\\
B^{(2)}_{r\phi}&=&-i\,e^{-{1\over 2Q}(t+y)}\,  
e^{i(\phi-kz)}\,\sin{\theta\over 2}\,\sin\theta\,
e^{-kr}\,{r^{1/2}\over 2}\nonumber\\
B^{(2)}_{\theta\phi}&=&-i\,e^{-{1\over 2Q}(t+y)}\,  
e^{i(\phi-kz)}\,\cos{\theta\over 2}\,\sin\theta\,
e^{-kr}\,{r^{3/2}\over 2}
\label{finalsol}
\eea
or in form language
\bea
B^{(2)}&=&e^{-{1\over 2Q}(t+y)}\,  
e^{i(\phi-kz)}\,e^{-kr}\,r^{1/2}\,\Bigl\{-
{1\over 2Q}\,\cos{\theta\over 2}\,dt\wedge dz\nonumber\\
&&+{r\over 2(Q+r)^2}\,\cos{\theta\over 2}\,
[dy-Q(1+\cos\theta) d\phi]\wedge \Bigl[dt - {2Q+r\over Q}\,dz\Bigr]\nonumber\\
&&+i k\,\cos{\theta\over 2}\,dr\wedge dz+{1\over 2}\sin{\theta\over  
2}\,dr\wedge [d\theta-i\sin\theta d\phi]
-{i\over 2}\,r\,\cos{\theta\over 2}\sin\theta\,d\theta \wedge  
d\phi\Bigr\}\nonumber
\eea

\section{Relation to D0-D6 bound states}

We have added one unit of P to a D1-D5 bound state. For the moduli we  
have used  this is a `threshold bound' state; i.e., the mass of the  
bound state is the mass of the D1-D5 plus the mass of the P.

In \cite{witten} it was noted that a D0 brane is repelled by a D6  
brane,\footnote{A similar system was also studied in \cite{mpt}.} 
but if a suitable flux $F$ was turned on in the D6 worldvolume  
then the D0 and D6 will form a  bound state. In this section we find a  
relation between our  threshold bound state and the condition in  
\cite{witten} which separates the domain of bound states from unbound  
states in the D0-D6 system.

\subsection{The near ring limit}\label{thenear}

Strictly speaking, we have a threshold bound state between the entire  
ring shaped D1-D5 and the entire P wavefunction. The threshold nature  
of this state follows from the general supersymmetry relation between  
D1,D5,P charges. But we have seen that for large $k$ the P wavefunction  
is confined to the vicinity of the ring, so we expect to get threshold  
binding between a short straight segment of the ring (like that in  
Fig.1(c)) and the wavefunction carrying P that we found in this short  
segment approximation. We will take the further step of identifying the  
two ends of our ring segment; this will enable us to perform a  
T-duality in the $z$ direction.

Let us review the charges carried by this segment of the ring:

\medskip

(a) We have the KK monopole charge that can be measured by a $S^2$  
surrounding the tube; the nontrivially fibered circle of this KK is the  
$y$ circle, and the directions $T^4,z$ are `homogeneous directions', so  
they behave like directions along the `KK-brane'.

(b) We have the `true charge' D5 along  $T^4\times S^1$.

(c) We have  the `true charge' D1  along the $S^1$.

(d) We have `dipole momentum' along the ring direction $z$; we call  
this $P_z$.

(e) The `test quantum' that we seek to bind to this background is a  
unit of $P_y$  (momentum along  $S^1$).

\medskip

The dipole charges are created automatically by the  binding of the  
true charges, and so there are relations between the values of the true  
and dipole charges. To find these relations it is convenient to dualize  
the D5-D1 charges to NS1-P.  The near ring limit of NS1-P was discussed  
in some detail in \cite{supertubegms}; we reproduce some relevant  
details here.

The NS1 string carries the momentum P through transverse oscillations,  
described by a profile $\vec F(t-y)$. In Fig.2(a) we open up this  
multiwound NS1 to exhibit this vibration profile. Since the NS1 is  
wrapped many times on the $S^1$ we find that a short segment of this  
oscillating string looks like Fig.2(b). We can see that the winding  
along the direction $y$ is due to the `winding charge' of the NS1,  
while the slant along the $z$ direction is due to the `derivative of  
the transverse displacement'  ${d\vec F\over dy}$ which we take to  
point along the $z$ direction.

\begin{figure}[htbp]
\vskip -.7truecm
\begin{center}
\includegraphics[width=4.5in]{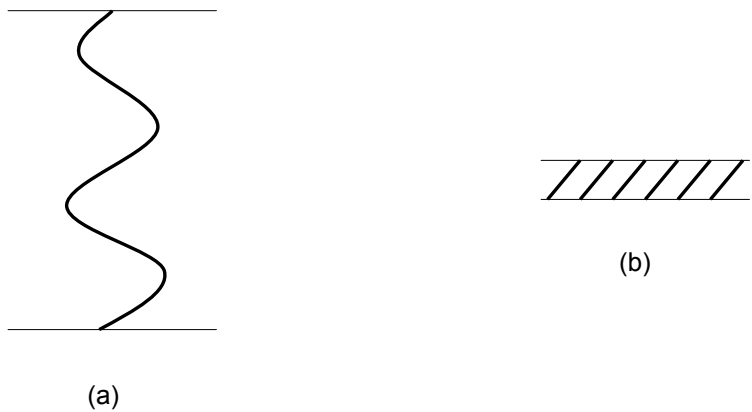}
\end{center}
\vskip -2.5truecm
\caption[x] {{\small  (a) The NS1 carrying a transverse oscillation profile in the
covering space of $S^1$. (b) The strands of the NS1 as they
appear in the actual space.}}
\label{fig1m}
\end{figure}

\begin{figure}[htbp]
\vskip -1truecm
\begin{center}
\hskip -2truecm
\includegraphics[width=4.0in]{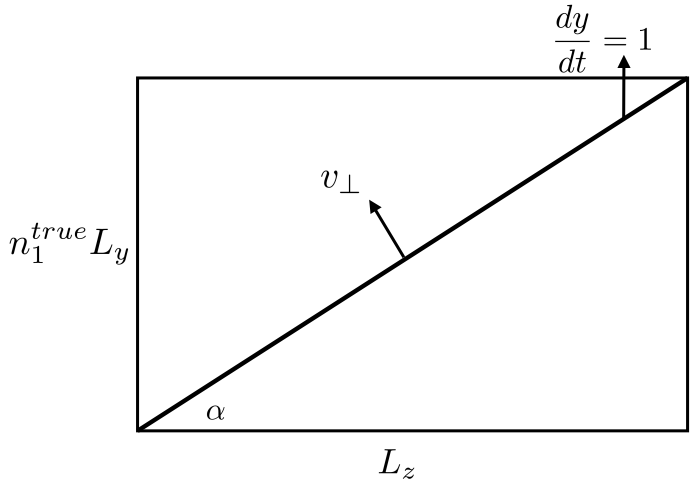}
\end{center}
\vskip -1.5truecm
\caption[x] {{\small The winding and momentum charges of a segment of the NS1; we have used  a  
multiple cover of the $S^1$ so that the NS1 looks like a diagonal line.}}
\label{fig2}
\end{figure}

For our present discussion we have compactified the $z$ direction, so  
that we can assign well defined `charges' to all elements.  We find the  
following charges in the NS1-P frame:

\medskip

(a) We have one unit of  $NS1_z$,  winding charge of the NS1 in the $z$  
direction. This is the dual of the KK dipole charge in the D1-D5 frame.  
Thus we write $n_1^{dipole}=1$.

(b) The `true' D5 charge becomes winding along $S^1$. We write this as  
$n_1^{true}$ units of $NS1_y$.

(c) The `true' D1 charge becomes momentum along $S^1$. We write this as  
$n_p^{true}$ units of $P_y$.

(d) We have momentum $P_z$ along the dipole direction (this has been  
unchanged in the duality from D1-D5).  The number of units of this  
momentum we call $n_p^{dipole}$.

(e) The quantum that we wish to bind to the background changes from  
$P_y$ to an NS5 along $S^1\times T^4$.

\medskip

Note that we must choose the compactification lengths of the $y,z$  
directions judiciously so that we get integer values for all charges.  
This can be done by choosing $L_z, L_y$ so that $n_1^{true},  
n_p^{true}$ are integers. We will see below (eq. (\ref{relationp}))  
that this will set $n_p^{dipole}$ to be integral. Note that  
$n_1^{dipole}=1$  so it is already integral.

\subsection{Relations between true and dipole charges}

{\bf Note on notation:}\quad In this section we will encounter three  
different duality related systems: D1-D5, NS1-P, and a system where  
these true charges become D4 branes. We will not need to compute in the  
D1-D5 frame. For the NS1-P frame we use unprimed symbols for all  
quantities (for example lengths are $L_y, L_z$ etc.). These should not  
be confused with unprimed symbols used in earlier sections of this  
paper; the computations here will not use results from those sections.  
For the frame using D4 branes we use primes on all symbols (e.g.   
$L'_y, L'_z$).

\bigskip

Let us ignore for now the charge (e) in the above list and look at the  
other charges which together give the background geometry of the ring.   
These charges are depicted in Fig.3. We have a NS1 moving in a  
direction perpendicular to itself with some velocity $v_\perp$; this  
gives all the four charges (a)--(d) above. We have denoted the lengths  
of the $y,z$ directions by $L_y, L_z$.

   We will now derive the relations between the true charges and the  
dipole charges. The first relation comes from the fact that the  
momentum carried by the NS1 is in a direction perpendicular to the NS1.  
Indeed if the momentum had a component along along the NS1 then there  
would be oscillations along the NS1 and a corresponding entropy.  The  
entire NS1-P bound state does has an entropy, which is manifested in  
different possible shapes for the entire ring. But we are now zooming  
in on a short segment of the ring, and so by definition should have no  
entropy visible in oscillations of this segment.

The NS1 winds in a direction given by the vector
\be
\vec W=L_z n_1^{dipole} \hat z+L_y n_1^{true} \hat y=L_z  \hat z+L_y  
n_1^{true} \hat y
\ee
The momentum vector is
\be
\vec P={2\pi n_p^{dipole}\over L_z }\hat z + {2\pi n_p^{true}\over L_y  
}\hat y
\ee
Requiring $\vec W\cdot \vec P=0$ gives
\be
n_1^{dipole}n_p^{dipole}+n_1^{true}n_p^{true}=0, ~~~\Rightarrow~~~  
n_p^{dipole}=-n_1^{true}n_p^{true}
\label{relationp}
\ee

The second condition comes from the fact that the waveform $\vec  
F(t-y)$ moves along the $y$ direction at the speed of light $v=1$.
   This implies that the velocity of the NS1 in the direction normal to  
itself is
\be
v_\perp=\cos\alpha={L_z\over \sqrt{(L_z)^2+(n_1^{true}L_y)^2}}
\ee
The mass of the NS1 is
\be
M=T\sqrt{(L_z)^2+(n_1^{true} L_y)^2}
\ee
where $T=1/2\pi\alpha'$ is the tension of the NS1.
 The momentum of the NS1 is in the direction normal to itself, and has  
magnitude
\be
|\vec P|=\sqrt{\Bigl({2\pi n_p^{dipole}\over L_z}\Bigr)^2+\Bigl({2\pi  
n_p^{true}\over L_y}\Bigr)^2 }
\ee
Setting $|\vec P|={Mv_\perp \over \sqrt{1-v_\perp^2}}$ we get
\be
\sqrt{\Bigl({2\pi n_p^{dipole}\over L_z}\Bigr)^2+\Bigl({2\pi  
n_p^{true}\over L_y}\Bigr)^2 }=
T\sqrt{(L_z)^2+(n_1^{true} L_y)^2}{L_z\over n_1^{true}L_y}
\label{relationpre}
\ee
Using (\ref{relationp}) gives
\be
\sqrt{\Bigl({2\pi n_p^{dipole}\over L_z}\Bigr)^2+\Bigl({2\pi  
n_p^{true}\over L_y}\Bigr)^2 }
={(2\pi) n_p^{true}\over L_yL_z}\sqrt{(L_z)^2+(n_1^{true} L_y)^2}
\ee
We thus find that  that (\ref{relationpre}) is equivalent to
\be
[Tn_1^{true}L_y]\Bigl[{2\pi n_p^{true}\over L_y}\Bigr]=[TL_z]^2
\label{condition5}
\ee
which tells us that
\be
[{\rm Mass~ of~ true ~NS1 ~charge}]\times [{\rm Mass ~of~ true ~P  
~charge}]~=~[{\rm Mass~ of~ NS1 ~dipole ~charge}]^2
\label{words}
\ee
In this form the condition is valid in all duality frames, with only  
the names of the charges changing under the dualities.

   To summarize we have two relations between the true and dipole  
charges. The relation (\ref{relationp})  comes from requiring that  
there be no entropy in the ring segment after we have zoomed into a  
sufficiently small region of the ring. The other condition  
(\ref{words})  is related to the supersymmetry of the charges  
distributed along the ring. The supersymmetry is assured by the fact  
that the entire waveform moves in one direction with the speed of  
light. Different parts of the NS1 have different slopes and different  
velocities $v_\perp$, but for a profile of the form $\vec F(t-y)$ the  
slope and velocity are always correlated in such a way that the  
different parts are mutually BPS.

   \subsection{Dualizing to D6-D0}

   We now wish to perform dualities that will map the dipole charge of  
the ring (KK in the case of D1-D5, NS1 in the case of NS1-P) to a D6  
brane charge. The quantum carrying one unit of $P_y$  will be converted  
to a D0. Since we have found the relations between true and dipole  
charges in the NS1-P frame let us start with NS1-P and perform the  
required dualities:
   \bea
\left(\begin{array}{c} NS1_z \\ NS1_y \\ P_y \\  P_z\\NS5_{y1234}  
\end{array} \right)
\stackrel{S}{\longrightarrow}\left(\begin{array}{c} D1_z \\ D1_y \\ P_y  
\\  P_z\\D5_{y1234} \end{array}
\right)
\stackrel{T_{yz12}}{\longrightarrow}\left(\begin{array}{c}  
\overline{D3}_{y12} \\ D3_{z12} \\ NS1_y \\  NS1_z\\\overline{D3}_{z34}  
\end{array} \right)\stackrel{S}{\longrightarrow}\left(\begin{array}{c}
\overline{D3}_{y12} \\  D3_{z12} \\ D1_y \\ D1_z\\ \overline{D3}_{z34}  
\end{array}  
\right)\stackrel{T_{z34}}{\longrightarrow}\left(\begin{array}{c}   
D6_{zy1234} \\ D4_{1234} \\ D4_{yz34} \\ D2_{34} \\D0\end{array}  
\right)\nn
\label{chain}
\eea
The true charges $n_1^{true}, n_p^{true}$ have become D4 branes which  
can be described by fluxes in the D6:
\bea
n_1^{true}&=&n_4^{(1234)}={1\over 2\pi}\int_{zy} F={L'_zL'_y\over 2\pi}  
F_{zy}\nn
n_p^{true}&=&n_4^{(yz34)}=-{1\over 2\pi}\int_{12} F=-{L'_1L'_2\over  
2\pi} F_{12}
\label{condition3}
\eea
where $L'_i$ are the lengths of cycles after the dualities. The minus  
sign in the expression for
$n_p^{true}$ arises from the orientation of the D6: the positive  
orientation is $(zy1234)$ while the $n_4$ is oriented as $(yz34)$.    
The presence of the above components of $F$ also induces a D2 charge
\be
n_2^{(34)}={1\over 2}{1\over (2\pi)^2}\int_{zy12} F\wedge  
F={L'_zL'_yL'_1L'_2\over (2\pi)^2}F_{zy}F_{12}= -
n_4^{(1234)}n_4^{(zy34)}
\label{relation}
\ee
Since under the dualities $n_p^{dipole}=n_2^{(34)}$, we observe that  
(\ref{relation}) is equivalent to (\ref{relationp}).
In other words the relation (\ref{relationp}) translates in the D6  
duality frame to the statement that the D2 charge comes entirely from  
the fluxes needed to induce the required D4 charges; there is no  
`additional' D2 charge.

\subsection{The condition of \cite{witten}}

Consider a D6 brane along the directions $(zy1234)$. Suppose that there  
is a background NS-NS 2-form turned on; by a suitable change of  
coordinates we can bring this to a form where the nonzero components  
are $b_1=B_{zy}, b_2=B_{12}, b_3=B_{34}$. Write
\be
e^{2\pi i v_a}={1+i b_a\over 1-i b_a}\quad a=1,2,3
\ee
The threshold value of $B$, beyond which  a D0 will bind to the D6, is  
given by \cite{witten}
\be
v_1+v_2+v_3={1\over 2}
\ee
In terms of the $b_a$ this condition becomes
\be
b_1 b_2 + b_1 b_3 + b_2 b_3 =1
\label{condition2}
\ee
We can replace the $B$ field with a field strength on the D6:
\be
b_a\to 2\pi\alpha' F_a
\ee
We take $\alpha'=1$ in the following. In our case we have $b_3=0$ and  
$b_1 = 2\pi\,F_{zy}$,
$b_2= 2\pi F_{12}$. One has the freedom to change the orientation in  
each of the 2-planes $(z,y)$, $(1,2)$,
$(3,4)$: this flips the sign of $v_a$ and $b_a$, and thus the sign of  
each of the terms in
(\ref{condition2}) is actually arbitrary. Taking this into account the  
threshold condition of \cite{witten} for our case is
\be
(2\pi)^2 |F_{zy} F_{12}|=1
\label{witten}
\ee

\subsection{Checking the threshold condition}

Let us now see if the condition (\ref{witten}) is satisfied by our ring  
segment. From (\ref{condition3}) we find that
\be
(2\pi)^2 F_{zy} F_{12}=-{(2\pi)^4\over  
L'_zL'_yL'_1L'_2}n_1^{true}n_p^{true}
\label{condition4}
\ee
Under the dualities (\ref{chain}) the moduli change as follows (primed  
quantities refer to D6 frame, unprimed to the NS1-P frame, and  
$L_i=2\pi R_i$)
\bea
\!\!\!\!\!\!\!\!\!\!
g'&=&g\,\sqrt{R_z\over R_y R_1 R_2}{1\over R_3 R_4}\,,\quad  
R'_3={g\over R_4\sqrt{R_z R_y R_1 R_2}}\,,\quad
R'_4={g\over R_3\sqrt{R_z R_y R_1 R_2}}\nonumber\\
\!\!\!\!\!\!\!\!\!\!R'_y &=& \sqrt{R_z R_1 R_2\over R_y}\,,\quad R'_z =  
\sqrt{R_z\over R_y  R_1 R_2}\,,\quad
R_1'=\sqrt{R_z R_y R_2\over R_1}\,,\quad R_2'=\sqrt{R_z R_y R_1\over  
R_2}
\label{moduli}
\eea
Thus
\be
(2\pi)^2 F_{zy} F_{12}=-{(2\pi)^4 n_1^{true} n_p^{true}\over  
L'_zL'_yL'_1L'_2}=-{(2\pi)^2n_1^{true}n_p^{true}\over L_z^2}
\ee
If we now use the relation (\ref{condition5}) we find
\be
(2\pi)^2 F_{zy} F_{12}=-1
\ee
We thus see that the charges carried by our ring satisfy the condition (\ref{witten}) noted in \cite{witten}.

\subsection{Depth of the tachyon potential}

Let us see what we have learned. The 2-charge system has true charges  
and dipole charges, and these satisfy the relations
(\ref{relationp}),(\ref{words}). The system can be mapped to a D6 brane  
carrying fluxes, and the fluxes have a value which puts the system at  
the boundary of the domain where a D0 brane will bind to the D6.

In the D1-D5 picture the analogue of the D0 is the P charge.  In  
section (\ref{thenear}) we listed the charges carried by the 2-charge  
D1-D5 system and the charge P carried by the wavefunction we are trying  
to construct. But there is one more charge carried by the wavefunction,  
which comes from the momentum of this wavefunction  along the $z$  
direction. In the wavefunction this momentum arises from the factor  
$e^{-ikz}$. So we would label this charge as an additional amount of  
$P_z$, carried by the quantum that we are trying to bind to the  
2-charge D1-D5 ring segment.

In the D6 duality frame this additional $P_z$ becomes a $D2_{34}$. Thus  
the quantum that we are trying to bind to the D6 is not just a D0, but  
a `D0 plus some $D2_{34}$'.  We now draw some conclusions about the  
D0-D6 bound state from  our construction of the wavefunction  
(\ref{finalsol}).

\subsubsection{The case $k=0$}

Since $k$ is a free parameter, we can try to set $k=0$. This would  
correspond to letting the test quantum be just the D0 (not bound to any $D2_{34}$), and asking if at the  
threshold value of fluxes (\ref{witten}) we get a good bound state with  
the D6. But from the discussion of section (\ref{norm}) we see that the  
wavefunction is {\it not} normalizable for the case $k=0$, so there is  
no  bound state in this case. We therefore conclude that for a D6  
wrapped on a  torus $T^6$ carrying fluxes at the threshold value  
(\ref{witten}) we do not get a bound state with the D0. As argued in  
\cite{witten} we would of course get a bound state for larger values of  
$F$ and no bound state for smaller $F$, but our explicit construction  
of the wavefunction (in the dual D1-D5 case) tells us the situation at  
the threshold value of $F$. 

\subsubsection{The case $k>0$}

In this case the test quantum to be bound has some $D2_{34}$ branes  
along with the D0. The mass of a
`D0 plus some $D2_{34}$' is obviously more than the mass of just the  
D0. But after we bind the `D0 plus some $D2_{34}$'
to the D6 carrying fluxes, the final mass of the bound state is  
independent of the mass of the $D2_{34}$ branes coming with the D0,  
since in the D1-D5 frame the energy of the wavefunction is given by  
$e^{-i{t\over 2Q}}$ for all values of $k$.

This observation tells us the binding energy of the  $D2_{34}$ branes  
in the situation where we have a D6 carrying fluxes equal to their  
`threshold' value (\ref{witten}). The binding energy $\Delta E$ must be  
equal to the mass $M_2$ of the $D2_{34}$ in order  that these branes do  
not show up in the final result for the mass of the composite:
\be
\Delta E=M_2
\label{bind}
\ee
Note that the D0 is repelled by a D6, is neutral with respect to the  
D4's in the D6, and is attracted by the D2 charge in the D6. At the  
`threshold' value of fluxes it becomes neutral with respect to the  
`D6-D4-D4-D2' bound state created by the D6 with fluxes. By contrast  
the $D2_{34}$ is neutral with respect to the D6, is attracted to the  
D4's in the D6, and is neutral with respect to the D2 in the D6. Thus  
we expect a binding energy $\Delta E$ for the $D2_{34}$, and our  
construction of the wavefunction tells us that this energy is  
(\ref{bind}). In CFT terms we get a tachyon in the open string spectrum  
between
the $D2_{34}$ and the D6 with fluxes. For the threshold value of these  
fluxes the depth of the tachyon potential must equal the mass of the  
$D2_{34}$.

\section{Discussion}

We have constructed a simple case of `3-charge hair' for the BPS black  
ring, by starting with a D1-D5 ring and adding a perturbation carrying  
one unit of P. A normalizable perturbation carrying this P was expected  
to exist because there was a corresponding state in the dual CFT. After  
constructing the perturbation we observe that it is smooth everywhere,  
so the result supports a `fuzzball' picture for black ring microstates.

A similar perturbation was constructed (up to several orders in a small  
parameter) for the 3-charge black {\it hole} in \cite{mss}, and  
solutions dual to specific CFT states carrying nonperturbative amounts  
of P  were found in \cite{3charge}. But these solutions carried a large  
amount of angular momentum. Thus it may be said that they did not give  
generic microstates for the 3-charge hole. By contrast, the black ring  
is {\it supposed} to carry a sizable amount of angular momentum, which  
gives it the `ring shape'. Thus even though we have only one unit of P  
in our present construction,  the hair we have  
constructed might be considered a good indicator of the nature of generic states of the ring.

In \cite{lm6} `ring-like' 2-charge states were considered, and it was observed that the area of a `horizon' drawn around such states has an area satisfying a Bekenstein type relation $A/4G\sim \sqrt{n_1n_5-J}\sim S$ where $S$ is the entropy of these states and $J$ is the angular momentum of the ring.  (Such 2-charge systems have been further studied recently \cite{is,bks,diis,adbm}.) In the present paper we have taken the simplest of the 2-charge ring states and added one unit of P. We have made the wavefunction only in the near ring limit, where the segment of the ring looked like a straight line. But we will get a similar near ring limit from any sufficiently smooth microstate out of the collection used in \cite{lm6}, so our wavefunction adding P should describe the nature of P excitations for any of these 2-charge microstates.

A large class of 3-charge BPS solutions for the black hole and  
black ring were found in \cite{bw2,gim}. While the explicit examples  
studied there had axial symmetry (and thus a nontrivial amount of  
rotation) one may be able to  construct nonrotating solutions by extending such techniques. 
Thus this approach may lead to generic nonperturbative hair for the black hole as  
well as for the black ring. It would therefore be very interesting to  
identify  microstates in this approach.  In the perturbative  
construction of the present paper we have excited the NS-NS 2-form  
gauge field, which was not excited in the solutions of \cite{bw2,gim}.  
It would be interesting to find an extension of the solutions of  
\cite{bw2,gim} which give nonperturbative hair involving this gauge  
field.

\section*{Acknowledgements}

  This work is
supported in part by DOE grant DE-FG02-91ER-40690. We thank Ashish Saxena for discussions.

\appendix

\section{Coordinates for the ring}

\renewcommand{\theequation}{A.\arabic{equation}}
\setcounter{equation}{0}

In this appendix we explain the geometric meaning of the coordinates  
(\ref{nrchange}) useful in describing the ring, and also obtain the  
near ring limit used in our analysis. The coordinates we define are  
constructed on the lines of the coordinates used in \cite{emparan}, and  
are related to them by a simple transformation.

The D1-D5 geometry (\ref{mm}) can be generated by starting with an  
NS1-P system where the NS1 describes one turn of a uniform helix. Let  
this helix lie in the $x_1-x_2$ plane of the noncompact 4-dimensional  
space $x_1,x_2,x_3,x_4$.  We introduce polar coordinates in this space
\bea
x_1=\t r\sin\t\theta\cos\t\phi, ~~~&x_2&=\t r\sin\t\theta \sin\t\phi\nn
x_3=\t r\cos\t\theta\cos\t\psi, ~~~&x_4&=\t r\cos\t\theta\sin\t\psi
\label{tilde}
\eea
Then the coordinates $\bar r,\bar \theta$ appearing in (\ref{mm}) are  
related to $\t r,\t\theta$ by \cite{lm4}
\be
\t r=\sqrt{\bar r^2+a^2\sin^2\bar\theta}, ~~~\cos\t\theta={\bar  
r\cos\bar\theta\over \sqrt{\bar r^2+a^2\sin^2\bar\theta}},
~~~\t\phi=\bar\phi, ~~~\t\psi=\bar\psi
\ee
In these coordinates the ring is easy to see; the center of the `tube'  
runs along the circle at $\t r=a, \t\theta=\pi/2$.  We will start by  
defining our ring coordinates with the help of these variables, and  
later convert to the coordinates $(\bar r,  
\bar\theta,\bar\phi,\bar\psi)$.

We want to define coordinates near the ring such that the direction  
along the ring becomes a linear coordinate
\be
z=a\bar\phi
\ee
We now wish to choose coordinates in the 3-dimensional space  
perpendicular to the ring.  Choose a point $P = (a\cos\t\phi,  
a\sin\t\phi, 0,0)$ on the ring. Close to the ring we would like these  
to be spherical polar coordinates $r,\theta,\phi$ centered at $P$, with  
the direction $\theta=0$ pointing towards the center of the ring. Close  
to the ring the coordinate $r$ should measure distance from the ring,  
but when $r\sim a$ we will see the diametrically opposite point $P' =  
(-a\cos\t\phi, -a\sin\t\phi, 0,0)$ on the ring, and should use a radial  
coordinate that vanishes at $P'$.  Consider all points that have  
azimuthal coordinate $\bar\phi=\tilde \phi$ and for these points define
\be
{1\over r}={1\over 2a}({r_P\over r_{P'}}+{r_{P'}\over r_P})
\ee
where $r_P, r_{P'}$ measure distances from the points $P$, $P'$ respectively
\be
r_P=\sqrt{\t r^2+a^2-2a\t r \sin\t\theta}, ~~~r_{P'}=\sqrt{\t  
r^2+a^2+2a\t r \sin\t\theta}
\ee
If we approach the point $P$ we have $r_P\r 0$, and
\be
{1\over r}\approx {1\over 2a}{r_{P'}\over r_P}\approx {1\over r_P}
\ee
So we see  that  $r\approx r_P$ near $P$, and similarly $r\approx r_{P'}$  
near $P'$.

Note that
\be
r_P^2r_{P'}^2=(a^2+\t r^2)^2-4\t r^2a^2\sin^2\t\theta=(a^2-\t r^2)^2+4  
\t r^2a^2\cos^2\t \theta
\ee
Thus
\be
|a^2-\t r^2|\le r_Pr_{P'}
\ee
with equality only for points on the ring diameter passing through  
$P$, $P'$. Thus we can define
\be
\cos\theta={(a^2-\t r^2)\over r_Pr_{P'}}
\label{cos}
\ee
Near $P$ we have
\be
r_P\approx r, ~~~r_{P'}\approx 2a, ~~~a^2-\t r^2=(a+\t r)(a-\t  
r)\approx 2a (a-\t r)
\ee
Close to $P$ we have
\be
\t r=\sqrt{(x_1^2+x_2^2)+(x_3^2+x_4^2)}\approx \sqrt{x_1^2+x_2^2}
\ee
where we have kept terms up to linear order in the displacement from $P$.  
Thus $a-\t r$ measures the distance $d$ from the $P$
along the diameter through $P$ (with $d$ positive for points inside the  
ring). We then see that
\be
\cos\theta\approx {(2a) d\over (2a) r}={d\over r}
\ee
and thus $\theta$ is the desired polar coordinate near $P$. Finally note  
that the $x_3-x_4$ plane  is perpendicular to the ring  and also to the  
diameter through $P$, so we define the azimuthal angle
\be
\phi=\tan^{-1}{x_4\over x_3}=\t\psi
\ee

Using (\ref{tilde}) we write the ring coordinates in terms of the  
coordinates   $(\bar r, \bar\theta,\bar\phi,\bar\psi)$
\be
r=a\big ( {\bar r^2+a^2\cos^2\bar\theta\over \bar  
r^2+a^2+a^2\sin^2\bar\theta}\big ),  
~~~\cos\theta={a^2\cos^2\bar\theta-\bar r^2\over  
a^2\cos^2\bar\theta+\bar r^2}, ~~~z=a\bar\phi, ~~~ \phi=\bar \psi
\ee
The inverse of these relations gives (\ref{nrchange})
\bea
&&\rb^2={a^2 r (1-\cos\theta)\over a+r\cos\theta}\,,\quad \sin^2\thetab  
= {a-r\over a+r\cos\theta}
\,,\quad \psib=\phi \,,\quad \phib={z\over a}
\eea

\end{document}